\documentclass[lettersize,journal]{IEEEtran}
\usepackage{amsmath,amsfonts}
\usepackage{algorithmic}
\usepackage{algorithm}
\usepackage{array}
\usepackage[caption=false,font=normalsize,labelfont=sf,textfont=sf]{subfig}
\usepackage{textcomp}
\usepackage{stfloats}
\usepackage{url}
\usepackage{verbatim}
\usepackage{graphicx}
\usepackage{float}
\usepackage{subfig}
\usepackage{cite}
\usepackage{gensymb}
\usepackage{multirow}
\usepackage{color}
\usepackage{bm}
\usepackage{pifont}
\usepackage{soul}
\usepackage{array}
\usepackage[hidelinks]{hyperref}
\usepackage[dvipsnames]{xcolor}
\usepackage{epstopdf}
\AppendGraphicsExtensions{.tif}

\begin{document}

\title{Geometric Distortion Guided Transformer for Omnidirectional Image Super-Resolution}

\author{Cuixin Yang, 
        Rongkang Dong, \IEEEmembership{Graduate Student Member,~IEEE,}
        Jun Xiao,
        Cong Zhang, \IEEEmembership{Member,~IEEE,} \\
        Kin-Man Lam, \IEEEmembership{Senior Member,~IEEE,}
        Fei Zhou,
        Guoping Qiu, \IEEEmembership{Senior Member,~IEEE}
\thanks{This work was supported by the Hong Kong Research Grants Council (RGC) Research Impact Fund (RIF) under Grant R5001-18.
\textit{(Corresponding authors: Cuixin Yang and Kin-Man Lam.)}

Cuixin Yang, Rongkang Dong, Jun Xiao, Cong Zhang, and Kin-Man Lam are with the Department of Electrical and Electronic Engineering, The Hong Kong Polytechnic University, Hong Kong (e-mail: cuixin.yang@connect.polyu.hk; rongkang97.dong@connect.polyu.hk; jun.xiao@connect.polyu.hk; cong-clarence.zhang@connect.polyu.hk; enkmlam@polyu.edu.hk).

Fei Zhou is with the College of Electronic and Information Engineering, Shenzhen University, Shenzhen 518060, China, and also with the Guangdong-Hong Kong Joint Laboratory for Big Data Imaging and Communication, Shenzhen 518060, China (email: flying.zhou@163.com).

Guoping Qiu is with the School of Computer Science, University of Nottingham, NG8 1BB Nottingham, United Kingdom, and Ningbo 315100, China (email: guoping.qiu@nottingham.ac.uk).

}
}


\IEEEpubid{\begin{minipage}{\textwidth}\ \centering
Copyright~\copyright~2024 IEEE. Personal use of this material is permitted. \\
However, permission to use this material for any other purposes must be obtained from the IEEE by sending an email to pubs-permissions@ieee.org.
\end{minipage}
}

\maketitle

\begin{abstract}
As virtual and augmented reality applications gain popularity, omnidirectional image (ODI) super-resolution has become increasingly important. Unlike 2D plain images that are formed on a plane, ODIs are projected onto spherical surfaces. Applying established image super-resolution methods to ODIs, therefore, requires performing equirectangular projection (ERP) to map the ODIs onto a plane. 
ODI super-resolution needs to take into account geometric distortion resulting from ERP.
However, without considering such geometric distortion of ERP images, previous methods only utilize a limited range of pixels and may easily miss self-similar textures for reconstruction.
In this paper, we introduce a novel Geometric Distortion Guided Transformer for Omnidirectional image Super-Resolution (GDGT-OSR). Specifically, a distortion modulated rectangle-window self-attention mechanism, integrated with deformable self-attention, is proposed to better perceive the distortion and thus involve more self-similar textures. Distortion modulation is achieved through a newly devised distortion guidance generator that produces guidance for the rectangular windows by exploiting the variability of distortion across latitudes. 
Furthermore, we propose a dynamic feature aggregation scheme to adaptively fuse the features from different self-attention modules.
We present extensive experimental results on public datasets and show that the new GDGT-OSR outperforms methods in existing literature. 

\end{abstract}

\begin{IEEEkeywords}
Omnidirectional image, Super-resolution, Distortion, Rectangle-window, Transformer.
\end{IEEEkeywords}

\section{Introduction} 
\label{sec:introduction}

\IEEEPARstart{o}{mnidirectional} imaging, also known as 360\degree\ imaging, is a fundamental technology for developing immersive virtual reality (VR) and augmented reality (AR) applications. 
In practice, omnidirectional images (ODIs) are viewed through head-mounted display devices, which means that the viewport will have a limited range. To visualize the details of a scene from a narrow field-of-view, the images need to be of very high resolutions. However, camera systems for capturing high-resolution ODIs are expensive, as are the costs of storing and transmitting high-resolution ODIs \cite{deng2021lau}. 

One way to tackle this issue is through image super-resolution (SR), which reconstructs a high-resolution (HR) image from a low-resolution (LR) input \cite{liu2020photo,wu2023sfhn,huang2024deep}. 
For convenience in storage and transmission, raw ODIs are generally projected into 2D planar representations. Equirectangular projection (ERP) is the most common method for representing ODIs \cite{azevedo2019visual}. 
\IEEEpubidadjcol

\begin{figure}[!t]
\centering
\includegraphics[width=0.48\textwidth]{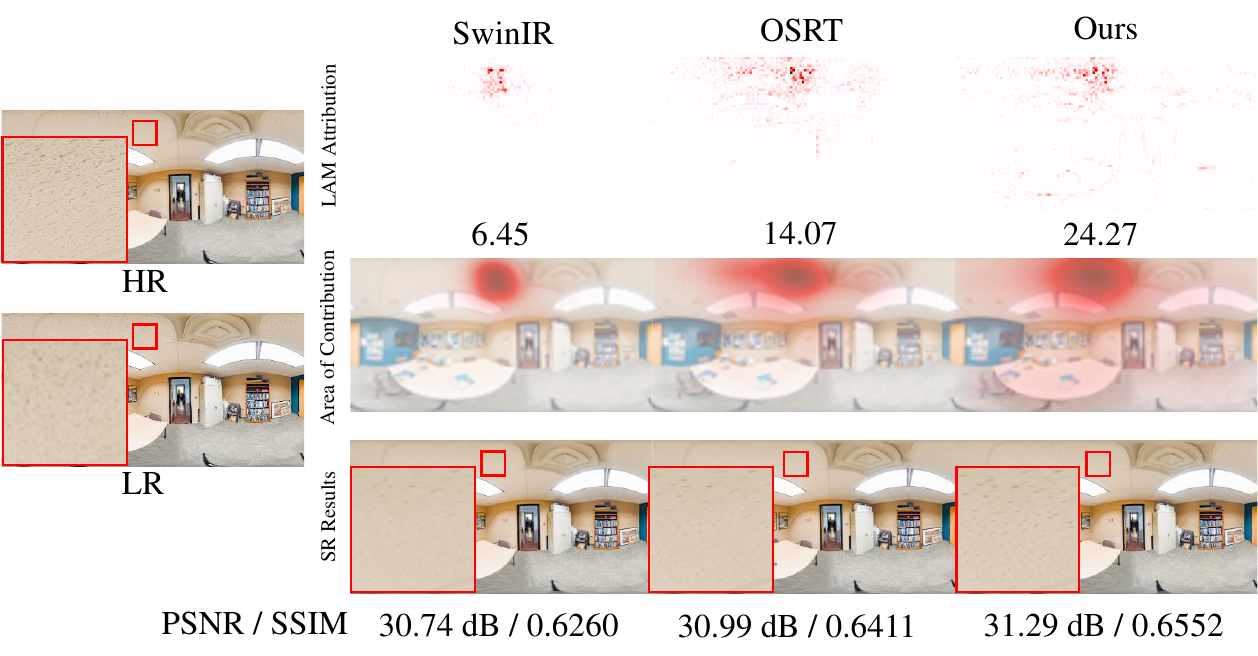}
\caption{Comparison of local attribution maps \cite{gu2021interpreting} and SR results among different methods. The local attribution maps represent the importance of each pixel in reconstructing the patch in the red box. The Diffusion Index (DI) is shown below the local attribution maps. A higher DI value indicates a wider range of the involved pixels. The second row shows the Area of Contribution, which implies the areas involved and their contributions. The local attribution maps, DI values, and Area of Contribution collectively demonstrate that our proposed method engages more pixels in the reconstruction. This contributes to restoring more realistic details, leading to improved SR performance.}
\label{sample}
\end{figure}

Transformer has emerged as a powerful and versatile computational paradigm \cite{dosovitskiy2020image}.
Local square-window self-attention is proposed to reduce the computational complexity of the global self-attention mechanism in vision transformer, leading to a limitation in the receptive field \cite{liang2021swinir,chen2022cross}.
However, for omnidirectional image super-resolution (ODISR), the ERP expands the ODIs and introduces distortion into ERP images. For example, a circle in the ODI is distorted into an oval in the ERP image. Small squared windows struggle to capture the whole oval, while large squared windows may encompass irrelevant patterns.
Therefore, square-window self-attention is a suboptimal option for reconstructing ERP images. Rectangular windows \cite{chen2022cross} can calibrate and expand the receptive field by involving more self-similar textures along the direction of stretching distortion, which is more appropriate for modeling features of ERP images than squared windows. However, existing ODISR methods \cite{deng2021lau,yoon2022spheresr,yu2023osrt} fail to consider extracting features of distorted ERP images from the perspective of the window's shape, leading to the limitation of involved pixels and self-similar textures. To solve this issue, we propose a Geometric Distortion Guided Transformer for Omnidirectional image Super-Resolution (GDGT-OSR), which aggregates features from windows of diverse shapes to calibrate and expand the attention area, involving more self-similar textures. Specifically, we propose a Distortion Modulated Rectangle-window Self-Attention (DMRSA) mechanism, which takes into account more self-similar regions of ERP images within a rectangular window. We integrate a deformable self-attention mechanism, which considers irregular neighborhoods and captures out-of-window similar patterns \cite{zhu2020deformable}, with DMRSA. Self-similarity is essential because it significantly contributes to the reconstruction of HR images \cite{zhou2014nonlocal,huang2015single,glasner2009super,michaeli2013nonparametric}. 
In DMRSA, Rectangle-window Self-Attention (Rwin-SA) is modulated by the distortion guidance that is generated through a newly devised Distortion Guidance Generator (DGG). DGG transforms the geometric distortion into distortion guidance, making the DMRSA adapt to the stretched ERP images. 
Furthermore, we dynamically aggregate the features from two self-attention modules by exploiting the information of windows with various shapes in Dynamic Feature Aggregation (DFA). 
To reveal the effects on the range of the involved area, we resort to Local Attribution Map (LAM) \cite{gu2021interpreting}, which is an attribution method for analyzing and visualizing attribution in SR.
As shown in Fig. \ref{sample}, our proposed method can utilize a wider range of self-similar information and more pixels to recover the patch in the red box, achieving a higher Diffusion Index (DI) and better SR results.

In summary, our main contributions are as follows:
\begin{itemize}

    \item We propose a novel framework, GDGT-OSR, designed for omnidirectional image super-resolution. By leveraging the distortion as guidance, GDGT-OSR effectively captures self-similar textures with windows of different shapes, leading to enhanced SR performance.
    \item To leverage the distortion information, we propose a Distortion Modulated Rectangle-window Self-Attention (DMRSA) mechanism paired with a Distortion Guidance Generator (DGG). The DGG transforms geometric distortion into distortion guidance, empowering DMRSA to adapt to the inherent distortion in ERP images through the newly designed distortion-guided rectangular windows.
    \item We devise a Dynamic Feature Aggregation (DFA) module to aggregate and complement features from different self-attention mechanisms based on the differences between them.
    \item Our GDGT-OSR framework achieves superior performance in ODISR, outperforming other state-of-the-art methods, including 2D plain SR and ODISR methods, in terms of quantitative and qualitative results.
    
\end{itemize}

The remaining parts of this paper are organized as follows. 
In Section \ref{sec:relatedwork}, we briefly review some related works. Section \ref{sec:methodology} introduces preliminaries about the related knowledge, followed by a detailed introduction of the proposed GDGT-OSR framework.
Section \ref{sec:experiments} presents the experimental settings, experiment results, and analysis of ablation studies. Finally, Section \ref{sec:conclusion} provides a summary, including the limitations and a brief conclusion of this paper.

\section{Related Works}
\label{sec:relatedwork}
In this section, we briefly review related works, including single image super-resolution (SISR), omnidirectional image super-resolution (ODISR), and Vision Transformer (ViT).
\subsection{Single Image Super-Resolution}
Long before the emergence of deep learning, researchers developed learning-based image resolution enhancement methods using image pyramids \cite{QIU1999} and look-up tables \cite{QIU2000360, freeman2002example, LI2009312}. After deep learning models demonstrated powerful capabilities in many related applications, researchers began to apply deep learning models, such as convolutional neural networks (CNNs), to image SR \cite{dong2015image,liu2020photo,wu2023sfhn,huang2024deep}. As recent literature is dominated by deep-learning-based methods, we review three types of popular deep-learning-based SR methods, including GAN-based SR, Transformer-based SR, and Diffusion-based SR. 

Generative Adversarial Networks (GANs) \cite{goodfellow2020generative} have shown powerful generative modeling abilities in many computer vision tasks, including SR \cite{ledig2017photo,wang2018esrgan,chan2021glean,chan2022glean}. SRGAN \cite{ledig2017photo} introduces an adversarial loss for GAN training to overcome the constraints associated with the PSNR-focused image SR. 
GLEAN \cite{chan2021glean, chan2022glean} uses pre-trained GANs, such as StyleGAN \cite{karras2019style} and BigGAN \cite{brock2019large}, as a latent bank to exploit their diverse priors. 

Since 2017, Transformers \cite{vaswani2017attention} have been widely used in computer vision. 
Researchers have applied Transformers to SR and demonstrated their efficacy in low-level computer vision tasks \cite{chen2021pre, liang2021swinir, chen2022cross, chen2023hat, chen2024recursive}.
IPT \cite{chen2021pre} introduces
a backbone model based on the standard Transformer to address diverse restoration challenges. Based on Swin Transformer \cite{liu2021swin}, SwinIR \cite{liang2021swinir} is proposed to solve different image restoration problems and has become a strong and popular backbone architecture for image restoration. However, using local squared windows, SwinIR suffers from a lack of direct interaction among windows, restricting the potential of establishing long-range dependencies. To solve this issue, CAT \cite{chen2022cross} proposes the Rwin-SA mechanism to broaden the attention area and increase the interaction across various windows by utilizing horizontal and vertical rectangle window attention. 

Recently, diffusion models\cite{sohl2015deep, ho2020denoising, song2021score, wu2024seesr, wang2024sinsr} have achieved unprecedented success in image/video synthesis and restoration. Due to the high computational cost resulting from pixel-space operation, LDM \cite{rombach2022high} suggests training diffusion models in the latent space to save computational resources. Since there is a strong connection between the LR inputs and the ground-truths when generating the outputs of SR, DiffIR \cite{xia2023diffir} only employs the diffusion model to estimate a compact image restoration prior representation with much fewer iterations than traditional diffusion models. Another way to make the diffusion model efficient is to increase the speed of inference\cite{yue2024resshift}.

The above methods focus on SR for 2D plain images, which are inappropriate for ODIs. In this paper, we propose a novel geometric distortion guided framework for ODISR.

\subsection{Omnidirectional Image Super-Resolution}
Initially, researchers began tackling the issue of SR for ODIs through spherical assembling \cite{nagahara2000super, arican2011joint, bagnato2010plenoptic}. These methods attempted to super-resolve an HR image from a series of consecutive LR ODIs under diverse projections. Subsequently, researchers shifted their research focus and began to study ODISR on plain images, i.e., equirectangular panorama images. Meanwhile, deep learning was introduced into ODISR by training the deep-learning-based model SRCNN \cite{dong2015image} and adapting it to equirectangular panorama images \cite{fakour2018360}. As GANs become popular in the computer vision community, they have been employed in ODISR \cite{ozcinar2019super, zhang2020toward}. In the work of \cite{ozcinar2019super}, the authors introduce a fast PatchGAN discriminator trained with a loss function designed for spherical images. In \cite{zhang2020toward}, an efficient multi-frequency GAN architecture is proposed to solve the SR of real-world panoramic images. However, these methods only address the distribution discrepancy between plain images and panoramic images and only fine-tune the plain image SR methods.

Deng et al. \cite{deng2021lau} pointed out that the pixel density is non-uniform and varies across latitudes in ERP projected ODIs. Therefore, they propose a progressive pyramid network, namely LAU-Net, to super-resolve the pixels at different latitude bands hierarchically, rather than making the model adapt to ODIs. However, training multiple levels of networks for different latitude bands is computationally expensive, and it can also lead to inconsistencies between latitude bands.
The work \cite{nishiyama2021360} directly concatenates a distortion map and an LR image as the input, which is fed into a network designed for 2D plain images. However, the information of the distortion map is not exploited sufficiently, and it is difficult for a 2D-image SR network to learn to restore ODIs without modification. SphereSR \cite{yoon2022spheresr} generates a continuous spherical image representation and employs the local implicit image function (LIIF) \cite{chen2021learning} to predict the RGB values under various projection types continuously. Although SphereSR can resolve ODIs with arbitrary projection types flexibly, it requires training multiple network branches for different projection types. 
TCCL-Net\cite{chai2023tccl} proposes a Transformer and convolution collaborative learning network for ODISR, which extracts both long-range and short-range dependencies in an end-to-end manner. Furthermore, in the work \cite{chai2023super}, the same authors further explore the SR of panoramic images from the left-right view by combining binocular information and panoramic characteristics.
When generating training pairs, all of these ODISR methods apply uniform bicubic downsampling on the ERP images, neglecting the geometric properties of ERP in the degradation process. OSRT \cite{yu2023osrt} utilizes Fisheye downsampling that applies uniform bicubic downsampling on the original ODIs. Even though OSRT makes use of the deformable Transformer to leverage the distortion information, the attention area is still small and inefficient for reconstructing ODIs and it is difficult to adaptively calibrate the features across latitudes. 
To solve this issue, we propose to enlarge the attention area by involving different kinds of self-attention with diverse window shapes. Furthermore, we introduce a distortion guidance generator to modulate the features based on the variability of distortion across latitudes.

\subsection{Vision Transformer}
Transformer \cite{vaswani2017attention} was first introduced to process sequences in natural language processing. It has been developed for various computer vision tasks and has achieved extraordinary performances, including image recognition \cite{dosovitskiy2020image, touvron2021training}, object detection \cite{carion2020end, zhu2020deformable}, and segmentation \cite{wang2021pyramid, zheng2021rethinking}. Due to their great potential in dealing with high-level computer vision tasks, Transformers have been adopted in the field of image restoration, such as IPT \cite{chen2021pre} and SwinIR \cite{liang2021swinir}. Uformer \cite{wang2022uformer} applies self-attention in an 8$\times$8 local windows and adopts the U-net architecture to capture both local and global dependencies. However, the local windows used in SwinIR and Uformer limit the range of long-range dependency representation and the receptive field of the Transformer model. CAT \cite{chen2022cross} adopts the Rwin-SA to enlarge the attention area. Furthermore, deformable attention is exploited in ViTs \cite{xia2022vision, yu2023osrt} to capture the content in irregular neighborhoods. 

The contents are non-uniformly stretched in ERP images. 
The squared windows, which are commonly utilized for regular 2D images, restrict the attention area from accommodating the distortion caused by the stretch in ERP images.
To address this limitation, we propose a distortion modulated Rwin-SA that incorporates deformable self-attention. By considering the shape and direction of distortion, this approach calibrates the attention area, enabling it to capture more similar textures and patterns in ERP images.

\section{Methodology}
\label{sec:methodology}

\subsection{Preliminaries}
In this section, we introduce the stretching ratio and distortion map for a better understanding of the projection relationship between ODIs and the projection plane, as well as the derivation of the distortion map.
\subsubsection{\textbf{Stretching Ratio}} 
\begin{figure}[!t]
\centering
\includegraphics[width=0.48\textwidth]{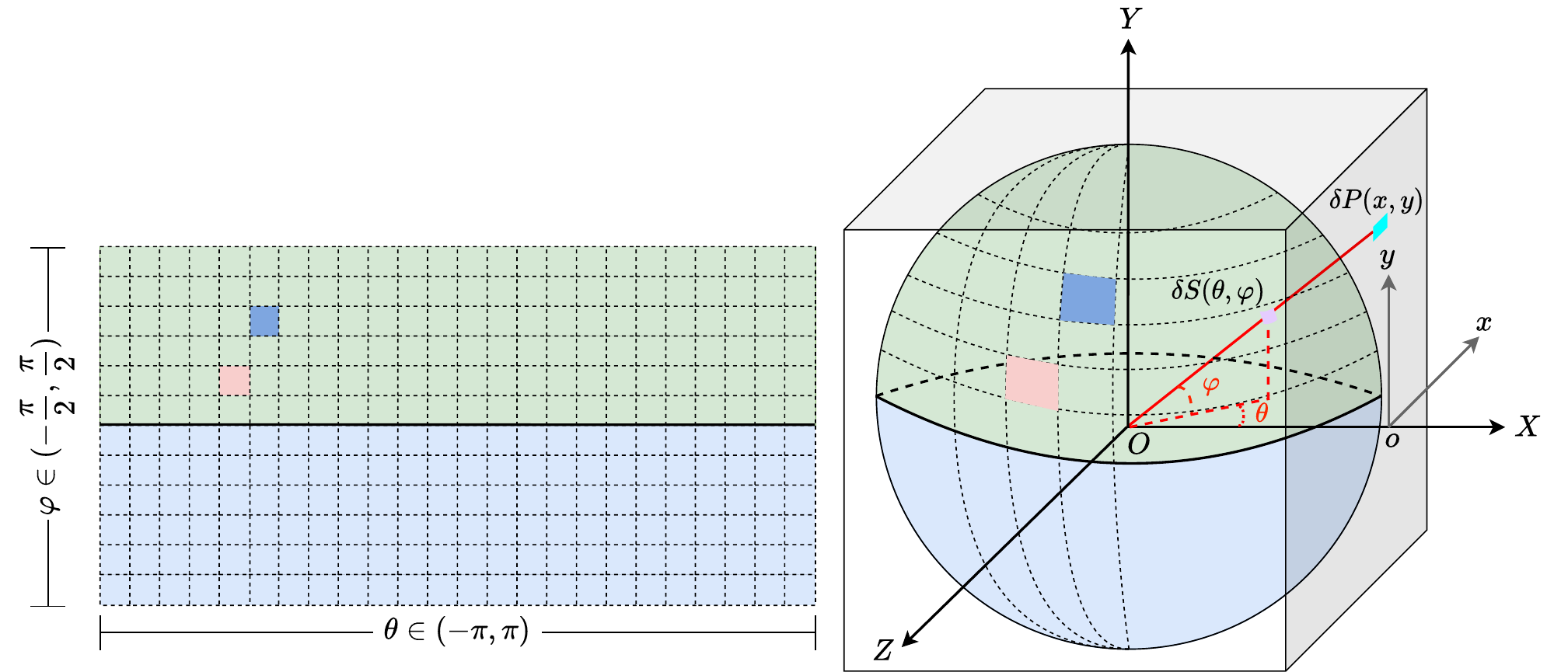}
\caption{Geometric explanation of the relationship between ERP (left) and the sphere, as well as the relationship between the sphere and the tangential cube (right).}
\label{geometric_explanation}
\end{figure}

As shown in Fig.~\ref{geometric_explanation}, each coordinate $(\theta, \varphi)$ on the ideal spherical surface corresponds to a point $(x, y)$ on the projection plane. The relationship between $(\theta, \varphi)$ and $(x, y)$ can be formulated as follows:
\begin{equation}
    x = h(\theta, \varphi), y = t(\theta, \varphi),
\label{equa_x_y}
\end{equation}
where $h(\cdot)$ and $t(\cdot)$ are coordinate transformation functions from the spherical surface to the projection plane.
Given a microunit $\delta S(\theta, \varphi)$ centered at $(\theta, \varphi)$ on the spherical surface and its corresponding microunit $\delta P(x, y)$ centered at $(x, y)$ on the projection plane, the stretching ratio is defined as follows:
\begin{equation}
    R(x, y) = \frac{\delta{S(\theta, \varphi)}}{\delta{P(x, y)}}=\frac{\cos(\varphi)|d{\theta}d{\varphi}|}{|dxdy|}=\frac{\cos(\varphi)}{|J(\theta, \varphi)|},
\label{R}
\end{equation}
where $J(\theta, \varphi)$ is a Jacobian determinant. Considering Eq.~(\ref{equa_x_y}), the Jacobian determinant is defined as follows:
\begin{equation}
J(\theta, \varphi)=\frac{\partial (x, y)}{\partial (\theta, \varphi)}=
\begin{vmatrix}
\frac{\partial x}{\partial \theta} & \frac{\partial x}{\partial \varphi} \\
\frac{\partial y}{\partial \theta} & \frac{\partial y}{\partial \varphi} \\
\end{vmatrix}.
\label{J}
\end{equation}
The projection relationship for ERP, which is a commonly used projection method for ODIs, is defined as follows \cite{sun2017weighted}:
\begin{equation}
    x= h(\theta, \varphi) = \theta, y = t(\theta, \varphi) = \varphi.
\label{x_y}
\end{equation}
Thus, derived from Eqs.~(\ref{R})-(\ref{x_y}), the stretching ratio of ERP can be formulated as follows:
\begin{equation}
    R_{ERP}(x, y) = \cos(\varphi) = \cos(y),
\label{R_ERP}
\end{equation}
where $x\in{(-\pi, \pi)}$, $y\in{(-\frac{\pi}{2}, \frac{\pi}{2}})$.

\begin{figure}[!t]
\centering
\includegraphics[width=2.2in]{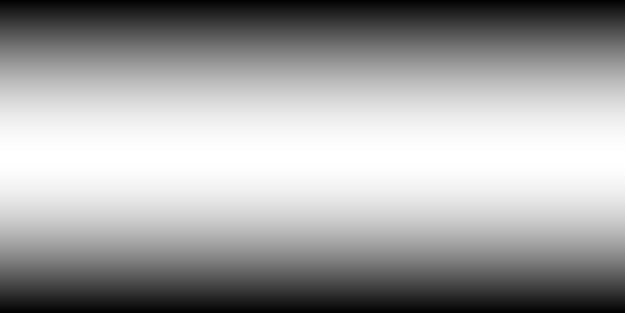}
\caption{Distortion map. A lighter area represents less distortion, while a darker area represents higher distortion.}
\label{D_map}
\end{figure}
\subsubsection{\textbf{Distortion Map}} The projection from raw ODIs to ERP images leads to distortion in the latter. This distortion varies along the latitude and is symmetric in the two hemispheres. According to Eq.~(\ref{R_ERP}), given an LR image $I^{LR}\in{\mathcal{R}^{{H}\times{W}\times{C_{in}}}}$ ($H$, $W$, and $C_{in}$ represent the height, width, and number of channels, respectively), the corresponding distortion map $D\in{\mathcal{R}^{{H}\times{W}\times{1}}}$ is defined as follows~\cite{sun2017weighted,yu2023osrt}:
\begin{equation}
\label{dist_map}
D(h, 1:W) = \text{cos}\frac{(h+0.5-H/2)\pi}{H},
\end{equation}
where $D(h, 1:W)$ represents the stretching ratio from an ideal spherical surface to the 2D ERP image at the current height of $h$.
As shown in Fig.~\ref{D_map}, the distortion around the equator area is the smallest, while the distortion intensifies as the latitude increases.

\subsection{Architecture}

Unlike regular 2D images, ERP images are non-uniformly stretched in the projection space. Local squared windows in traditional Transformers make it difficult for the attention area to adapt to the stretched distortion. 
An attention area that takes this distortion into account is more appropriate for distorted ERP images. 
To address this issue, we propose a geometric distortion guided framework called GDGT-OSR, which includes Distortion Modulated Rectangle-window Self-Attention (DMRSA) and Distortion-aware Deformable Self-Attention (DDSA). 
By considering the distortion characteristics of ERP images, GDGT-OSR calibrates and expands the attention area. This allows it to capture more similar textures and patterns, which are crucial for SR reconstruction.

\begin{figure*}[!t]
\centering
\includegraphics[width=6.8in]{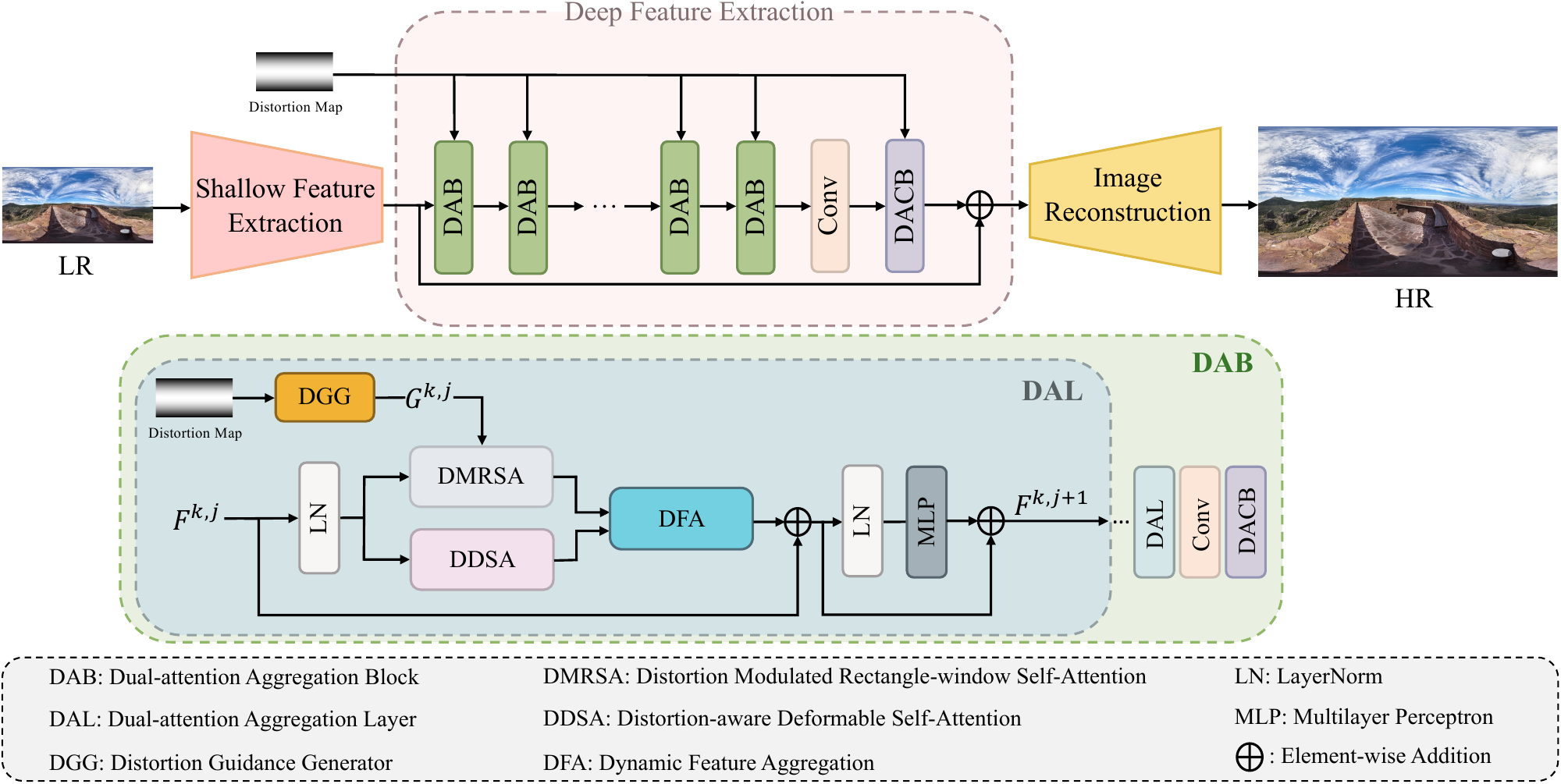}
\caption{Overview of the GDGT-OSR architecture (upper part) and the detailed structure of the DAB (bottom part).}
\label{overview}
\end{figure*}

\subsubsection{\textbf{Overview}}
An overview of the proposed method is shown in Fig.~\ref{overview}. It consists of three modules: shallow feature extraction, deep feature extraction, and image reconstruction. Our model takes an LR ODI $I^{LR}\in{\mathcal{R}^{{H}\times{W}\times{C_{in}}}}$ and the corresponding distortion map $D\in{\mathcal{R}^{{H}\times{W}\times{1}}}$ as input and reconstructs an HR ODI. We utilize only one convolutional layer in the shallow feature extraction module to obtain the low-level feature $F^{0}\in{\mathcal{R}^{{H}\times{W}\times{C}}}$ ($C$ is the number of channels). This feature is then fed into the deep feature extraction module. This module comprises $K$ Dual-attention Aggregation Blocks (DABs), each of which includes several Dual-attention Aggregation Layers (DALs), a convolutional layer, and a Distortion-Aware Convolution Block (DACB) \cite{yu2023osrt}. 

The structure of the proposed DAB is depicted at the bottom of Fig.~\ref{overview}. The Distortion Modulated Rectangle-window Self-Attention (DMRSA) and Distortion-aware Deformable Self-Attention (DDSA) mechanisms generate features characterized by different
local windows. Specifically, in DMRSA, we leverage the proposed Distortion Guidance Generator (DGG) to 
adaptively modulate the features according to the latitudes. Furthermore, for better feature fusion, features from DMRSA and DDSA are dynamically aggregated within a Dynamic Feature Aggregation (DFA) module according to their respective importance. 

\subsubsection{\textbf{Distortion Guidance Generator (DGG)}} The DGG is designed to extract a representative feature, i.e., the distortion guidance, that encapsulates the distortion. This distortion guidance is utilized to adaptively modulate the key and value features from the Rwin-SA to better adapt to the unevenly stretched content in ERP images. Thus, the output feature of the self-attention mechanism is calibrated by the distortion guidance. The process of DGG can be expressed as follows:
\begin{equation}
{G}=\text{DGG}\left(D\right),
\end{equation}
where $D$ and $G$ denote the distortion map and the distortion guidance, respectively. 

\begin{figure*}[!t]
\centering
\includegraphics[width=0.9\textwidth]{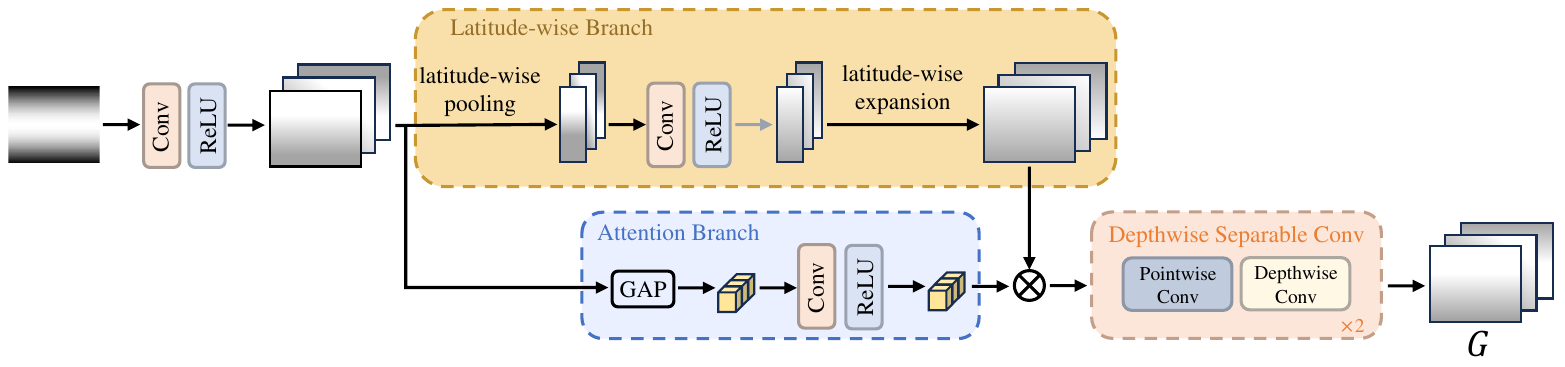}
\caption{Illustration of the Distortion Guidance Generator (DGG).}
\label{DoM}
\end{figure*}

From the geometric property of the distortion map, it can be noted that the distortion for locations at the same latitude is identical. Based on this fundamental property, we propose to model the distortion map in a latitude-wise manner. Specifically, the distortion map $D\in{\mathcal{R}^{{H}\times{W}\times{1}}}$ is first convolved to extract feature maps $f\in{\mathcal{R}^{{H}\times{W}\times{C}}}$. Then, in the Latitude-wise branch, each feature map undergoes latitude-wise pooling (LWP). In LWP, as shown in Fig.~\ref{DoM}, the pixel values of each feature map are averaged by row, transforming a feature map into a vector. The process is expressed as follows: 
\begin{equation}
\label{LWP}
v^{i}=\text{LWP}(f^{i}),
\end{equation}
\begin{equation}
\label{AVG}
{\rm where\ } v^{i}_{j}=\frac{1}{W} \sum_{w=1}^{W}\left(f^{i}_{j,w}\right), j\in[1,H]. 
\end{equation}
In Eq.~(\ref{LWP}) and (\ref{AVG}), $v^{i}$ and $f^{i}$ denote the $i$-th vector and feature map, respectively, $v^{i}_{j}$ represents the $j$-th element of the $i$-th vector, and $f^{i}_{j,w}$ represents the $w$-th element of the $j$-th row in the $i$-th feature map.
After LWP, the column vectors are then convolved by a 1$\times$1 convolutional layer followed by an activation layer, as follows: 
\begin{equation}
\label{conv+relu}
v^{\prime}=\text{ReLU}(\text{Conv}(v)).
\end{equation}
Latitude-wise expansion (LWE) is then applied to the learned vectors $v^{\prime}$. In latitude-wise expansion, each element in the column vector is duplicated $W$ times to restore the size $(H\times{W})$ of the original feature map.
\begin{equation}
\label{eq:lwe}
g^{i}=\text{LWE}((v^{\prime})^{i}), 
\end{equation}
\begin{equation}
\label{eq:lwe2}
{\rm where \ } g^{i}_{j,1:W}=(v^{\prime})^{i}_{j}, j\in[1,H]. 
\end{equation}
In Eq.~(\ref{eq:lwe}) and (\ref{eq:lwe2}), $g^{i}$ denotes the $i$-th feature map of the expanded features $g$ generated by LWE, $g^{i}_{j,1:W}$ denotes all the elements of the $j$-th row in the $i$-th feature map of $g$.

The other branch, i.e., the attention branch, aims to calculate the attention to adaptively weight the expanded features from latitude-wise expansion based on the distortion feature maps $f$. Specifically, the attention branch consists of a Global Average Pooling (GAP) layer, a convolutional layer, followed by an activation layer. The expanded features from the latitude-wise branch are multiplied by the learned attention weights. In this way, the learned attention weights adaptively weigh the expanded features. This process can be formulated as follows:
\begin{equation}
\label{weighted_g}
\hat{g}^{i}=w^{i}\cdot{g^{i}}, 
\end{equation}
where $w$ represents the learned attention weights, $i$ represents the $i$-th element, and $\hat{g}$ denotes the distortion guidance. 

Depthwise separable convolution \cite{chollet2017xception} is then adopted to refine $\hat{g}$, which consists of a depthwise convolutional layer and a pointwise convolutional layer. The advantage of utilizing the depthwise separable convolution in the refinement is three-fold. Firstly, for a specific latitude, its distortion is similar to that of its neighboring latitudes. A depthwise convolutional layer can leverage the features of local neighboring latitudes to complement and enrich the features at the current latitude. Secondly, a pointwise convolutional layer fuses the features of different channels at the same location, which means that the features from different channels are integrated pixel by pixel and latitude by latitude. Thirdly, the computational cost is reduced by adopting depthwise separable convolution, compared to traditional convolution.

The advantages of the proposed DGG are two-fold. Firstly, DGG can encode the distortion map in latent space and utilize the distortion in an implicit way. Secondly, DGG can leverage the distortion prior to effectively guide the expansion of the attention area. We will demonstrate the effectiveness of DGG in Section~\ref{subsec:ablation}.

\subsubsection{\textbf{Distortion Modulated Rectangle-Window Self-Attention (DMRSA)}} 
\begin{figure*}[!t]
\captionsetup[subfigure]{font=scriptsize,labelfont=footnotesize}
\centering
\subfloat[Rectangle Window Self-Attention]{\includegraphics[width=0.43\textwidth]{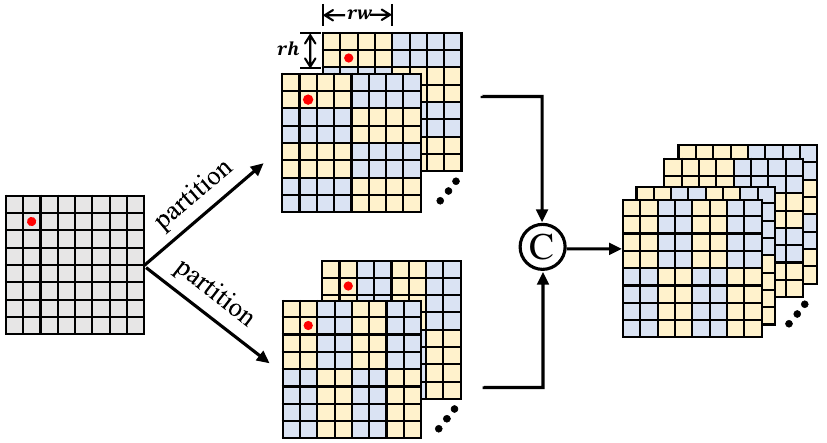}} \hspace{1cm}
\subfloat[Distortion Modulated Rectangle Window Self-Attention]{\includegraphics[width=0.43\textwidth]{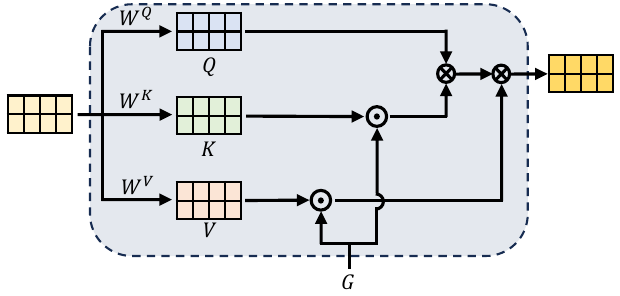}}
\caption{(a) Illustration of the rectangle window self-attention scheme. (b) The overall structure of the distortion modulated rectangle-window self-attention. The key $K$ and value $V$ features are modulated by $G$ learned from the distortion map by the DGG module. $W^{Q}$, $W^{K}$, $W^{V}$ are transformation matrices for $Q$, $K$, $V$.}
\label{MRWS}
\end{figure*}

As mentioned in Section \ref{sec:introduction}, the ERP images are non-uniformly stretched in the projection.
Specifically, Fig.~\ref{D_map} shows the stretching ratio of a whole ERP image. 
From Eq.~(\ref{R}), we understand that the stretching ratio represents the ratio between the area of a microunit on the ideal spherical surface and that on the projection plane. Therefore, it can be inferred that at high latitudes, a small area on the ideal spherical surface is mapped to a large area on the projection plane, while at low latitudes, the distortion is smaller but still exists.
Considering this geometric distortion, we propose to adopt the rectangle window self-attention scheme, which can also expand the attention area by providing more informative textures and details \cite{chen2022cross}. Moreover, based on the intrinsic geometric distortion, we leverage the proposed DGG generator to provide modulations for the key and value features according to the distortion of different latitudes.

\textbf{Rectangle Window Self-Attention (Rwin-SA).} The rectangle window self-attention scheme is illustrated in Fig.~\ref{MRWS}(a). If $rh < rw$, the rectangle windows are horizontal windows, denoted as H-Rwins. Conversely, if $rh > rw$, the rectangle windows are vertical windows, denoted as V-Rwins. For a given input $X\in\mathcal{R}^{H\times{W}\times{C}}$, where $C$ is the number of channels, it is divided into a number of $rh\times{rw}$ non-overlapping H-Rwins or V-Rwins in each attention head. 
Specifically, given $N$ attention heads, where $N$ is even, these attention heads are equally divided into two parts. The outputs of H-Rwins and V-Rwins are concatenated along the channel dimension. The process for the $i$-th rectangle window can be expressed as follows:
\begin{equation}
\label{yi}
Y^{i}=\text{Concat}\left(Y^{i}_{1}, Y^{i}_{2}, ..., Y^{i}_{N}\right),
\end{equation}
where $Y^{i}_{1}, ..., Y^{i}_{\frac{N}{2}}$ are the outputs of the $\frac{N}{2}$ attention heads using H-Rwin, while $Y^{i}_{\frac{N}{2}+1}, ..., Y^{i}_{N}$ are the outputs of the remaining attention heads using V-Rwin.

\textbf{Distortion Guided Rwin-SA.} In ODIs, the pixel density is non-uniform across latitudes due to latitude-wise distortions. As depicted in Fig.~\ref{D_map}, the pixel density exhibits the highest level of compactness, with the distortion being mildest around the equator area. Conversely, in the high-latitude areas, particularly in the polar area, the pixel density appears considerably sparser, accompanied by a noticeable distortion.
The distorted pixels with different distortions across latitudes should contribute to the reconstruction of the current patch differently. We argue that the neighboring latitude-wise pixels with similar distortions are more related to the current patch because of the projection, which should be paid more attention to and contribute more.
Therefore, it is necessary to leverage the geometric distortion information of ERP images for reconstructing a high-quality image.
As shown in Fig.~\ref{MRWS}(b), the output of the DGG generator $G$ is utilized
to modulate the key and value features of the Rwin-SA based on the distortion map. 
The element-wise dot product is conducted between the output modulation features and the key features, as well as the value features, which can be expressed as follows:
\begin{equation}
\label{m_value}
\tilde{K}^{i}_{n}=K^{i}_{n}\odot{G}, \tilde{V}^{i}_{n}=V^{i}_{n}\odot{G},
\end{equation}
where $\tilde{K}^{i}_{n}$ and $\tilde{V}^{i}_{n}$ denote the modulated key and value features, respectively, $G$ is the distortion guidance, and $\odot$ represents the element-wise multiplication. 
Note that $D$, i.e., the distortion map, is shared across different heads. With query $Q^{i}_{n}$, the modulated key $\tilde{K}^{i}_{n}$ and value $\tilde{V}^{i}_{n}$, DMRSA is formulated as follows:
\begin{gather}
\label{m_x}
\tilde{Y}^{i}_{n}=\text{SoftMax}\left(\frac{Q^{i}_{n}(\tilde{K}^{i}_{n})^{T}}{\sqrt{d}}+B\right){\tilde{V}^{i}_{n}},\\
\tilde{Y}^{i}=\text{Concat}\left(\tilde{Y}^{i}_{1}, \tilde{Y}^{i}_{2}, ..., \tilde{Y}^{i}_{N}\right),\\
\text{DMRSA}(X)=\left(\tilde{Y}^{1}, \tilde{Y}^{2}, ..., \tilde{Y}^{\frac{H\times{W}}{rh\times{rw}}}\right){W^{f}},
\end{gather}
where $B$ represents dynamic relative position encoding \cite{wang2023crossformer++}, $d$ is the channel dimension of each head, $\tilde{Y}^{i}$ denotes the modulated output of the $i$-th window, and $W^{f}$ is the projection matrix for feature aggregation. In DMRSA, the key and value features are adaptively modulated by geometric distortion across latitudes.

\begin{figure}[!t]
\centering
\includegraphics[width=0.48\textwidth]{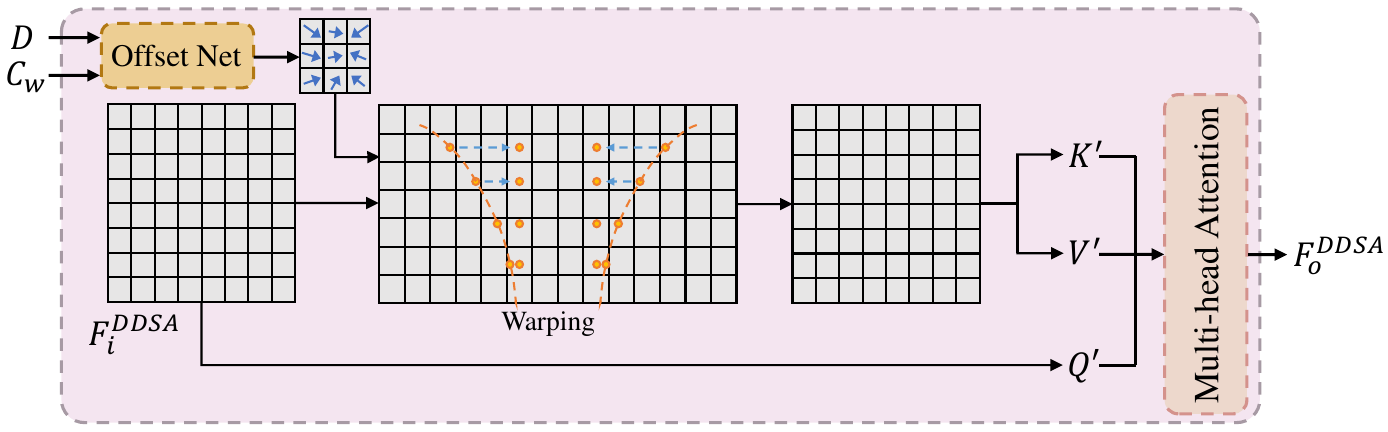}
\caption{Illustration of Distortion-aware Deformable Self-Attention (DDSA).}
\label{DDSA}
\end{figure}

DMRSA can calibrate and expand the attention area along the direction of stretching distortion, involving most related and self-similar textures and patterns. For those similar and out-of-window textures, we adopt the Distortion-aware Deformable Self-Attention (DDSA) mechanism\cite{yu2023osrt} as shown in Fig.~\ref{DDSA}, which considers irregular neighborhoods and provides a distortion-dependent attention area for flexibly modeling features, to complement DMRSA in our DAL.

\subsubsection{\textbf{Dynamic Feature Aggregation (DFA)}}
\begin{figure}[!t]
\centering
\includegraphics[width=0.45\textwidth]{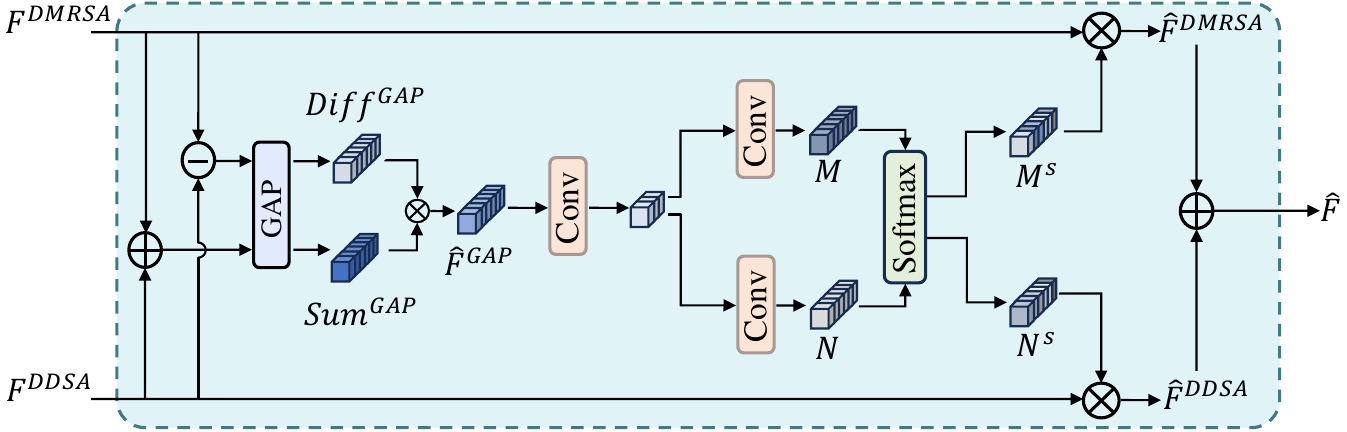}
\caption{Illustration of Dynamic Feature Aggregation (DFA).}
\label{DFA}
\end{figure}

Inspired by \cite{li2019selective}, we propose to dynamically aggregate the features from DMRSA and DDSA in a DFA module to better fuse the features from the dual-attention mechanism. Fig.~\ref{DFA} illustrates the DFA scheme, where features from DMRSA ($F^{DMRSA}$) and DDSA ($F^{DDSA}$) are firstly summed element-wisely. 
The difference between $F^{DMRSA}$ and $F^{DDSA}$ reveals the distinction between DMRSA and DDSA. This distinction implies the features neglected by either DMRSA or DDSA, such as the out-of-window features neglected by DMRSA, to which we should pay more attention. Therefore, we propose to compute the difference between $F^{DMRSA}$ and $F^{DDSA}$. The difference, denoted as $Diff$, is then utilized to weigh the summed features. Specifically, the Global Average Pooling (GAP) is performed on the $Diff$ and summed features. Thus, two features, i.e.,
$Diff^{GAP}$ and $SUM^{GAP}$, which condense the information along the channel dimension, are obtained. The weighted feature $\hat{F}^{GAP}$ is calculated as follows:
\begin{equation}
\label{weighted_F}
\hat{F}^{\rm GAP} = Diff^{\rm GAP} \odot SUM^{\rm GAP}.
\end{equation}
After that, $\hat{F}^{GAP}$ is further condensed by reducing the dimension through a convolutional layer, and then the compact feature is convolved to generate attention vectors, i.e., $M$ and $N$, for $F^{DMRSA}$ and $F^{DDSA}$, respectively. Specifically, softmax is conducted on $M$ and $N$ as follows: 
\begin{equation}
\label{soft-attn}
M_{i}^{s}=\frac{e^{M_{i}}}{e^{M_{i}}+e^{N_{i}}},
N_{i}^{s}=\frac{e^{N_{i}}}{e^{M_{i}}+e^{N_{i}}},
\end{equation}
where $M_{i}$ and $N_{i}$ denote the $i$-th element of $M$ and $N$, respectively. $M_{i}^{s}$ and $N_{i}^{s}$ denote the $i$-th element of $M$ and $N$ after softmax, respectively. The original input features are weighed by the attention vectors:
\begin{equation}
\label{modulate}
\hat{F}^{DMRSA}={M^{s}} \cdot F^{DMRSA}, \hat{F}^{DDSA}={N^{s}} \cdot F^{DDSA}.
\end{equation}
Finally, we obtain the output of DFA by summing them, which can be expressed as follows:
\begin{equation}
    \hat{F} = \hat{F}^{DMRSA} + \hat{F}^{DDSA}.
\end{equation} 
DFA takes into account the distinction between the features from different self-attention mechanisms, which should be addressed in feature fusion. By complementing these features, DFA learns to adjust them adaptively.

\subsection{Loss Function}
Due to the non-uniform pixel densities, it is unreasonable to calculate the pixel-wise errors in ERP images using the standard $l1$ or $l2$ loss. The pixel-wise errors at different locations in the ERP image should have different significance to reflect such uneven distribution. 
Therefore, we adopt the Weighted-to-Spherically $l1$ loss (WS-$l1$ loss \cite{baniya2023omnidirectional}) as the loss function in the training process. The WS-$l1$ loss can calibrate the pixel-wise errors based on the distortion map as follows:
\begin{equation}
\label{ws-l1}
\text{WS-}l1=\sum_{h=1}^{H}\sum_{w=1}^{W}|I^{gt}(h,w)-I^{o}(h,w)|D(h,w),
\end{equation}
where $I^{gt}$ and $I^{o}$ denote the ground-truth
image and the output image, respectively. $D(h,w)$ represents the value of the distortion map at the coordinate $(h,w)$. 
From Eq.~(\ref{dist_map}), we can know that the weights are larger at lower latitudes, while the weights are smaller at higher latitudes. 
From the perspective of non-uniform pixel density, this design is reasonable: the pixel density is more compact at lower latitudes while it is sparser at higher latitudes, which means more pixels are distributed around lower latitudes, so these pixels should contribute more to the loss function, and vice versa.

\begin{table*}[ht]
\centering
\caption{Quantitative comparison of different methods in two public datasets, i.e., ODI-SR\cite{deng2021lau} and SUN 360 Panorama\cite{xiao2012recognizing}. The scaling factors are 2 and 4. The best results are highlighted in bold.}
\begin{tabular}{l|c|cccc|cccc}
\hline
\multirow{2}{*}{Method} & \multirow{2}{*}{Scale} & \multicolumn{4}{c|}{ODI-SR\cite{deng2021lau}}                                         & \multicolumn{4}{c}{SUN 360 Panorama\cite{xiao2012recognizing}}     \\
                        &                        & PSNR           & SSIM            & WS-PSNR        & WS-SSIM         & PSNR           & SSIM            & WS-PSNR        & WS-SSIM         \\ \hline\hline
Bicubic                 & \multirow{9}{*}{$\times{2}$}    & 28.21          & 0.8215          & 27.61          & 0.8156          & 28.14          & 0.8118          & 28.01          & 0.8321          \\
RCAN\cite{zhang2018image}                    &                        & 30.26         & 0.8777          & 29.61          & 0.8739          & 30.84          & 0.8793          & 31.39          & 0.9008          \\
SRResNet\cite{wang2018esrgan}                &                        & 30.12          & 0.8703          & 29.56          & 0.8685          & 30.57          & 0.8692          & 31.13          & 0.8937          \\
EDSR\cite{lim2017enhanced}                    &                        & 30.18          & 0.8740          & 29.57          & 0.8708          & 30.70          & 0.8743          & 31.24          & 0.8970          \\
SwinIR\cite{liang2021swinir}                  &                        & 30.64          & 0.8821          & 30.00          & 0.8777          & 31.33          & 0.8855          & 31.98          & 0.9059          \\
HAT\cite{chen2023activating,chen2023hat} & & 30.67 & 0.8821 & 30.05 & 0.8780 & 31.37 & 0.8858 & 32.06 & 0.9065 \\
RGT\cite{chen2024recursive} & & 30.46 & 0.8781 & 29.86 & 0.8753 & 31.10 & 0.8793 & 31.79 & 0.9023 \\
OSRT\cite{yu2023osrt}                    &                        & 30.77          & 0.8846          & 30.11          & 0.8795          & 31.52          & 0.8888          & 32.14          & 0.9081          \\
GDGT-OSR (ours)                    &                        & \textbf{30.87} & \textbf{0.8863} & \textbf{30.21} & \textbf{0.8811} & \textbf{31.67} & \textbf{0.8910} & \textbf{32.33} & \textbf{0.9099} \\ \hline\hline
Bicubic                 & \multirow{9}{*}{$\times{4}$}    & 25.59          & 0.7118          & 24.95          & 0.6923          & 25.29          & 0.6993          & 24.90          & 0.7083          \\
RCAN\cite{zhang2018image}                    &                        & 26.90          & 0.7618          & 26.21          & 0.7486          & 27.10          & 0.7649          & 27.01          & 0.7851          \\
SRResNet\cite{wang2018esrgan}                &                        & 26.91          & 0.7592          & 26.24          & 0.7447          & 27.10          & 0.7613          & 26.99          & 0.7802          \\
EDSR\cite{lim2017enhanced}                    &                        & 26.87          & 0.7612          & 26.18          & 0.7467          & 27.11          & 0.7643          & 26.97          & 0.7830          \\
SwinIR\cite{liang2021swinir}                  &                        & 27.31          & 0.7735          & 26.61          & 0.7589          & 27.71          & 0.7804          & 27.64          & 0.7996          \\
HAT\cite{chen2023activating,chen2023hat} & & 27.29 & 0.7717 & 26.61 & 0.7578 & 27.67 & 0.7783 & 27.60 & 0.7982 \\
RGT\cite{chen2024recursive} & & 27.31 & 0.7711 & 26.63 & 0.7571 & 27.69 & 0.7782 & 27.65 & 0.7982 \\
OSRT\cite{yu2023osrt}                    &                        & 27.41          & 0.7762          & 26.70          & 0.7609          & 27.84          & 0.7835          & 27.77          & 0.8020          \\
GDGT-OSR (ours)                    &                        & \textbf{27.44} & \textbf{0.7776} & \textbf{26.72} & \textbf{0.7623} & \textbf{27.92} & \textbf{0.7850} & \textbf{27.81} & \textbf{0.8036} \\ \hline 
\end{tabular}
\label{quan results}
\end{table*}

\begin{table*}[ht]
\centering
\caption{Quantitative comparison of different methods under large scaling factors, i.e., specifically $\times8$ and $\times16$. The best results are highlighted in bold.}
\begin{tabular}{l|c|cccc|cccc}
\hline
\multirow{2}{*}{Method} & \multirow{2}{*}{Scale} & \multicolumn{4}{c|}{ODI-SR\cite{deng2021lau}}                                         & \multicolumn{4}{c}{SUN 360 Panorama\cite{xiao2012recognizing}}       \\
                        &                        & PSNR           & SSIM            & WS-PSNR        & WS-SSIM         & PSNR           & SSIM            & WS-PSNR        & WS-SSIM         \\ \hline\hline
Bicubic                 & \multirow{6}{*}{$\times{8}$}    & 24.24          & 0.6525          & 23.51          & 0.6204          & 23.73          & 0.6369          & 23.22        & 0.6324              \\
SwinIR\cite{liang2021swinir}                  &                       & 25.09          & 0.6873          & 24.38          & 0.6619          & 24.97          & 0.6881          & 24.57         & 0.6953          \\
HAT\cite{chen2023activating,chen2023hat} & & 25.15 & 0.6909 & 24.42 & 0.6653 & 25.12 & 0.6936 & 24.69 & 0.7008 \\
RGT\cite{chen2024recursive} & & 24.86 & 0.6889 & 24.08 & 0.6632 & 24.68 & 0.6893 & 24.23 & 0.6963 \\
OSRT\cite{yu2023osrt}                    &                        & 25.28          & 0.6929          & 24.53          & 0.6664          & 25.32          & 0.6964          & 24.86          & 0.7027          \\
GDGT-OSR (ours)                    &                        & \textbf{25.32} & \textbf{0.6949} & \textbf{24.60} & \textbf{0.6687} & \textbf{25.42} & \textbf{0.6994} & \textbf{25.00} & \textbf{0.7068} \\ \hline\hline
Bicubic                 & \multirow{6}{*}{$\times{16}$}    & 22.66          & 0.6171          & 21.90          & 0.5785          & 22.04          & 0.6025          & 21.42          & 0.5899          \\
SwinIR\cite{liang2021swinir}                  &                        & 23.33          & 0.6388          & 22.58          & 0.6046          & 22.86          & 0.6292          & 22.28          & 0.6230          \\
HAT\cite{chen2023activating,chen2023hat} & & 23.32 & 0.6401 & 22.55 & 0.6057 & 22.87 & 0.6306 & 22.27 & 0.6246 \\
RGT\cite{chen2024recursive} & & 23.17 & 0.6391 & 22.40 & 0.6049 & 22.70 & 0.6297 & 22.10 & 0.6236 \\
OSRT\cite{yu2023osrt}                    &                        & 23.49          & 0.6421          & 22.71          & 0.6074          & 23.11          & 0.6339          & 22.48          & 0.6278          \\
GDGT-OSR (ours)                    &                        & \textbf{23.54} & \textbf{0.6425} & \textbf{22.78} & \textbf{0.6087} & \textbf{23.21} & \textbf{0.6356} & \textbf{22.60} & \textbf{0.6303} \\ \hline 
\end{tabular}
\label{large sf results}
\end{table*}

\section{Experiments}
\label{sec:experiments}
\subsection{Experimental Settings}
\textbf{Data and Evaluation.} 
In our experiments, we adopted the ODI-SR dataset\cite{deng2021lau}, which consists of 800 HR images, as our training dataset. The resolution of the HR ERP images is 1024$\times$2048. However, when training on such a small dataset, the model will easily suffer from overfitting. Therefore, in order to avoid the overfitting problem, we follow OSRT\cite{yu2023osrt} and include the DF2K-ERP dataset as a part of our training dataset. The DF2K-ERP dataset generates synthetic ERP images from 2D plain images in the DF2K dataset. The DF2K-ERP dataset consists of 146,000 HR ERP image patches with a patch size larger than 256$\times$256. The evaluation datasets include the ODI-SR testing dataset\cite{deng2021lau} and the SUN360 Panorama dataset\cite{xiao2012recognizing}, both of which contain 100 images. The resolution of the HR ERP images from these two different testing datasets is 1024$\times$2048. In our experiments, we implement scaling factors of 2$\times$ and 4$\times$ by downsampling the HR ERP images with the corresponding factors. Furthermore, following OSRT\cite{yu2023osrt}, fisheye downsampling is applied to the HR ERP images to generate LR-HR image pairs. Besides the commonly used PSNR and SSIM, WS-PSNR and WS-SSIM 
re-weighted with distortion are also adopted as our evaluation metrics.

\textbf{Implementation Details.}
We conducted our experiments with PyTorch\cite{paszke2019pytorch}. During the training stage, the batch size is 16, and the patch size is 64$\times$64. If there is no extra demonstration, the default $rh$ and $rw$ are 8 and 64 for H-Rwins, while for V-Rwins, the default $rh$ and $rw$ are 64 and 8. The sizes of $rh$ and $rw$ will be explored in the later Section~\ref{subsec:ablation}. The initial learning rate is set as $2\times{10^{-4}}$, and reduced by half at 250K, 400K, 450K, and 475K iterations. The total number of training iterations is 500K. The whole model consists of 6 DABs, and each DAB contains 6 DALs. The dimension of the embedding feature is 156. Adam \cite{kingma2015adam} is adopted as the optimizer in the training process, with $\beta_{1}=0.9$ and $\beta_{2}=0.99$.

\subsection{Comparisons with State-of-the-Art Methods}
\textbf{Quantitative Results.}
Table~\ref{quan results} shows the quantitative comparisons among different methods under $\times2$ and $\times4$ scaling factors. OSRT is an ODI-SR method, and we report the results as in \cite{yu2023osrt}. The remaining compared methods, e.g., RCAN \cite{zhang2018image}, SRResNet \cite{wang2018esrgan}, EDSR \cite{lim2017enhanced}, SwinIR \cite{liang2021swinir}, HAT \cite{chen2023hat} and RGT \cite{chen2024recursive} are SR methods for 2D plain images. Thus, we retrained them on the ODI-SR dataset plus the auxiliary DF2K-ERP dataset. 
RCAN, SRResNet, EDSR, and SwinIR have inferior performance to OSRT and our GDGT-OSR, which demonstrates that the key to ODISR is the adaptiveness of the distortion information.
With the dual-attention aggregation and distortion-guided components, our GDGT-OSR has strong distortion-awareness and distortion transformation abilities and achieves state-of-the-art (SOTA) performance on the two public datasets for both $\times2$ and $\times4$ tasks. This illustrates that our proposed method can better exploit and transform the distortion map to enhance the SR performance than OSRT.


To further evaluate the effectiveness of GDGT-OSR on larger scaling factors, we conducted experiments with scaling factors of $\times 8$ and $\times 16$. The results, presented in Table~\ref{large sf results}, compare GDGT-OSR with other SOTA SR methods, including SwinIR \cite{liang2021swinir}, HAT \cite{chen2023hat}, RGT \cite{chen2024recursive}, and OSRT \cite{yu2023osrt}. As shown in Table~\ref{large sf results}, despite the increased difficulty associated with larger scaling factors, our proposed GDGT-OSR consistently outperforms other SOTA 2D plain image SR and ODISR methods. This demonstrates the robustness and generalization capability of GDGT-OSR across different scaling factors, highlighting its effectiveness.

\textbf{Qualitative Results.}
\begin{figure*}[ht]
\centering
\includegraphics[width=\textwidth]{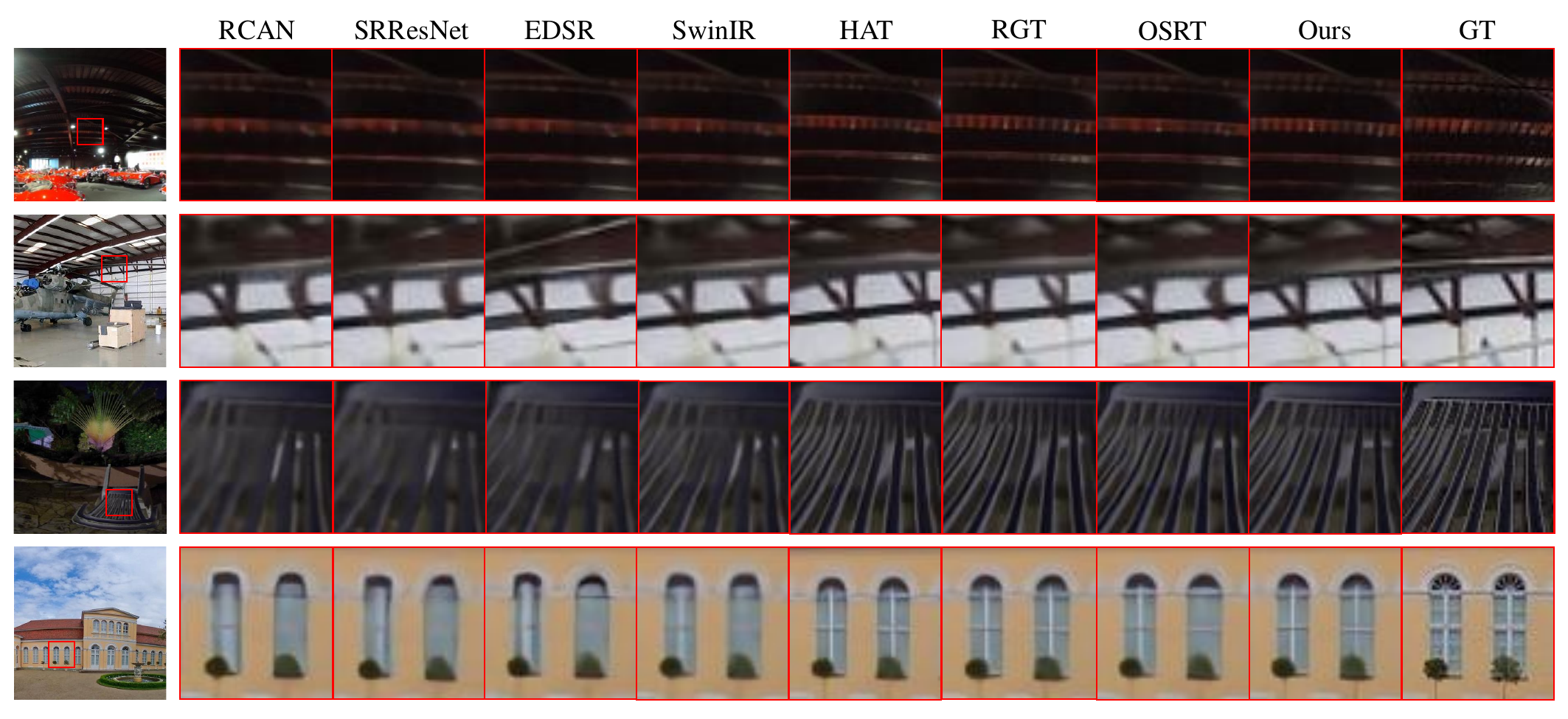}
\caption{Visual comparison of different methods on the ODI-SR dataset\cite{deng2021lau} with a scaling factor of 4. The red boxes highlight zoomed-in patches for better visualization.}
\label{sr_visual}
\end{figure*}
\begin{figure*}[ht]
\centering
\includegraphics[width=\textwidth]{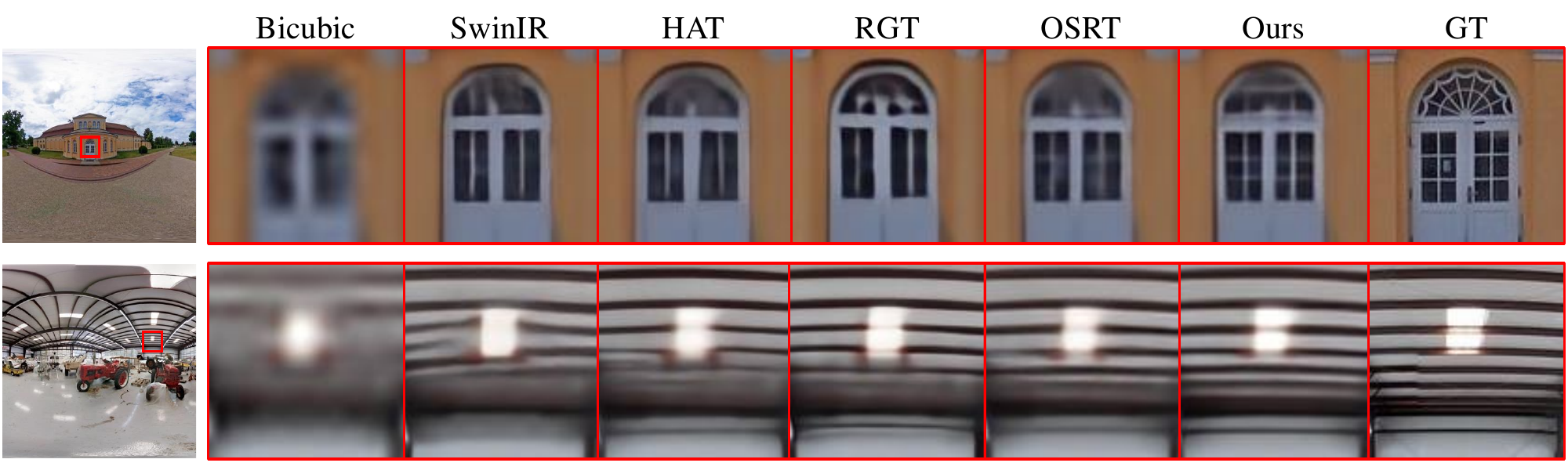}
\caption{Visual comparison of different methods on the ODI-SR dataset\cite{deng2021lau} with a scaling factor of 8. The red boxes highlight zoomed-in patches for better visualization.}
\label{sr_visual_x8}
\end{figure*}
Fig.~\ref{sr_visual} presents a visual comparison of different SR methods on the ODI-SR dataset with a scaling factor of 4. The red-boxed patches are zoomed in for better visualization.
Although the 2D plain image SR methods are retrained with the ODI datasets, their performances are still worse than those of OSRT and our GDGT-OSR. For example, as shown in the third sample in Fig.~\ref{sr_visual}, the ODISR methods, i.e., OSRT and our GDGT-OSR, can restore more visually pleasing results with less blurriness and distortion of the chair compared to the preceding four methods. This demonstrates that the inclusion and exploration of the distortion map contribute to the superiority and effectiveness of the last two ODISR methods.
Although OSRT represents SOTA performance in ODISR methods, our proposed method surpasses it by restoring more details and richer textures. For example, in the last sample in Fig.~\ref{sr_visual}, our method can restore comparatively complete window frames, while some parts of the window frames are missing in the visual result of OSRT. 

Fig.~\ref{sr_visual_x8} shows the visual comparison of different methods under the $\times8$ scaling factor. It is more difficult to super-resolve the images under a larger scaling factor because more details are lost in the LR images. As shown in Fig.~\ref{sr_visual_x8}, under the $\times8$ scaling factor, most of the SOTA SR methods, including SwinIR \cite{liang2021swinir}, HAT \cite{chen2023hat}, RGT \cite{chen2024recursive} and OSRT \cite{yu2023osrt}, struggle to recover the details and textures, leading to unsatisfactory visual results. With the proposed dual-attention and distortion-guided mechanisms, GDGT-OSR effectively manages distortion transformation and can recover more details for the ERP images even under the challenging large scaling factor.
To summarize, the above visual results demonstrate the superiority and effectiveness of the proposed GDGT-OSR under various scaling factors.

\begin{table*}[ht]
\centering
\caption{Ablation study of DMRSA and DDSA. The scaling factor is 4.}
\begin{tabular}{c|c|cccc|cccc}
\hline
\multirow{2}{*}{DMRSA} & \multirow{2}{*}{DDSA} & \multicolumn{4}{c|}{ODI-SR\cite{deng2021lau}}         & \multicolumn{4}{c}{SUN 360 Panorama\cite{xiao2012recognizing}}       \\
                      &                       & PSNR           & SSIM            & WS-PSNR        & WS-SSIM         & PSNR           & SSIM            & WS-PSNR        & WS-SSIM         \\  \hline\hline
\ding{55}             & \ding{55}                     & 27.10          & 0.7666          & 26.39          & 0.7503          & 27.46          & 0.7714          & 27.28          & 0.7883          \\
\ding{55}           & \ding{51}                     & 27.38          & 0.7753          & 26.66          & 0.7599          & 27.77          & 0.7819          & 27.70          & 0.8005          \\
\ding{51}             & \ding{55}                     & 27.39          & 0.7759          & 26.66          & 0.7605          & 27.83          & 0.7830          & 27.72          & 0.8015          \\
\ding{51}             & \ding{51}                     & \textbf{27.44} & \textbf{0.7776} & \textbf{26.72} & \textbf{0.7623} & \textbf{27.92} & \textbf{0.7850} & \textbf{27.81} & \textbf{0.8036} \\
\hline
\end{tabular}
\label{MRSA_DDSA}
\end{table*}

\begin{table*}[ht]
\centering
\caption{Ablation study of DGG and Diff. The scaling factors are 2 and 4.}
\begin{tabular}{c|c|c|cccc|cccc}
\hline
\multirow{2}{*}{Scale} & \multirow{2}{*}{DGG} & \multirow{2}{*}{Diff} & \multicolumn{4}{c|}{ODI-SR\cite{deng2021lau}}         & \multicolumn{4}{c}{SUN 360 Panorama\cite{xiao2012recognizing}} \\
                       &                      &                       & PSNR  & SSIM   & WS-PSNR & WS-SSIM & PSNR   & SSIM    & WS-PSNR & WS-SSIM \\ \hline\hline
\multirow{4}{*}{$\times{2}$}     & \ding{55} & \ding{55}  & 30.83 & \textbf{0.8863} & 30.17   & 0.8810  & 31.60  & 0.8908  & 32.25   & 0.9097  \\
                       & \ding{55}      & \ding{51}       & 30.84 & 0.8860 & 30.19   & 0.8809  & 31.63  & 0.8904  & 32.29   & 0.9095  \\
                       & \ding{51}           & \ding{55}  & \textbf{30.87} & \textbf{0.8863} & 30.20   & 0.8809  & \textbf{31.67}  & 0.8909  & 32.32   & 0.9098  \\
                       & \ding{51} & \ding{51}            & \textbf{30.87} & \textbf{0.8863} & \textbf{30.21}   & \textbf{0.8811}  & \textbf{31.67}  & \textbf{0.8910}  & \textbf{32.33}   & \textbf{0.9099}  \\ 
                       \hline\hline
\multirow{4}{*}{$\times{4}$}     & \ding{55} & \ding{55}  & 27.43 & 0.7775 & 26.71   & 0.7621  & 27.87  & 0.7847  & 27.78   & 0.8032  \\
                       & \ding{55}        & \ding{51}     & \textbf{27.44} & 0.7775 & \textbf{26.72}   & 0.7621  & 27.88  & 0.7848  & 27.79   & 0.8033  \\
                       & \ding{51}            & \ding{55} & 27.43 & \textbf{0.7779} & 26.71   & \textbf{0.7626}  & 27.89  & \textbf{0.7853}  & 27.80   & \textbf{0.8038}  \\
                       & \ding{51}  & \ding{51}           & \textbf{27.44} & 0.7776 & \textbf{26.72}   & 0.7623  & \textbf{27.92}  & 0.7850  & \textbf{27.81}   & 0.8036  \\ \hline 
\end{tabular}
\label{ablation study}
\end{table*}

\begin{table*}[!t]
\centering
\caption{Result comparison among variants of DGG. The scaling factor is 4.}
\begin{tabular}{c|cccc|cccc}
\hline
\multirow{2}{*}{Variants} & \multicolumn{4}{c|}{ODI-SR\cite{deng2021lau}}                                                                                      & \multicolumn{4}{c}{SUN 360 Panorama\cite{xiao2012recognizing}}                                                                            \\
                          & \multicolumn{1}{c}{PSNR} & \multicolumn{1}{c}{SSIM} & \multicolumn{1}{c}{WS-PSNR} & \multicolumn{1}{c|}{WS-SSIM} & \multicolumn{1}{c}{PSNR} & \multicolumn{1}{c}{SSIM} & \multicolumn{1}{c}{WS-PSNR} & \multicolumn{1}{c}{WS-SSIM} \\ \hline\hline
Direct                    & 27.42                    & 0.7769                   & 26.71                       & 0.7618                       & 27.86                    & 0.7841                   & 27.79                       & 0.8031                      \\ \hline
One conv                  & 27.43                    & 0.7765                   & 26.71                       & 0.7614                       & 27.88                    & 0.7841                   & 27.80                       & 0.8029                      \\ \hline
w/o atten                 & \textbf{27.44}           & 0.7771                   & \textbf{26.72}              & 0.7619                       & 27.88                    & 0.7843                   & 27.79                       & 0.8029                      \\ \hline
GDGT-OSR                   & \textbf{27.44}           & \textbf{0.7776}          & \textbf{26.72}              & \textbf{0.7623}              & \textbf{27.92}           & \textbf{0.7850}          & \textbf{27.81}              & \textbf{0.8036}             \\ \hline
\end{tabular}
\label{varients of DoM}
\end{table*}

\subsection{Ablation Study}
\label{subsec:ablation}

\begin{figure}[!t]
\captionsetup[subfigure]{font=tiny,labelfont=scriptsize}
\centering
\subfloat[Polar Area]{\includegraphics[width=0.45\textwidth]{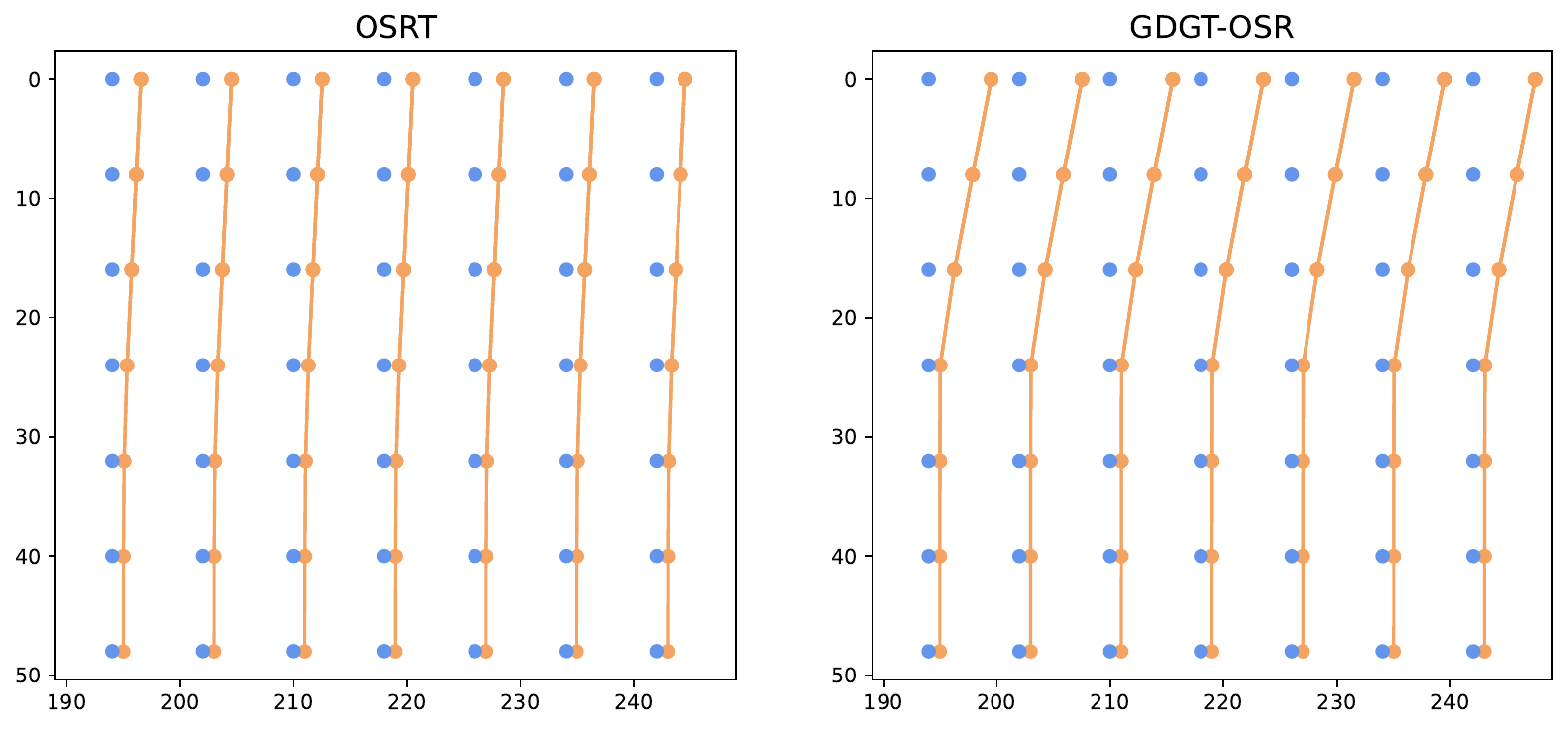}}\vskip -1pt

\subfloat[Equator Area]{\includegraphics[width=0.45\textwidth]{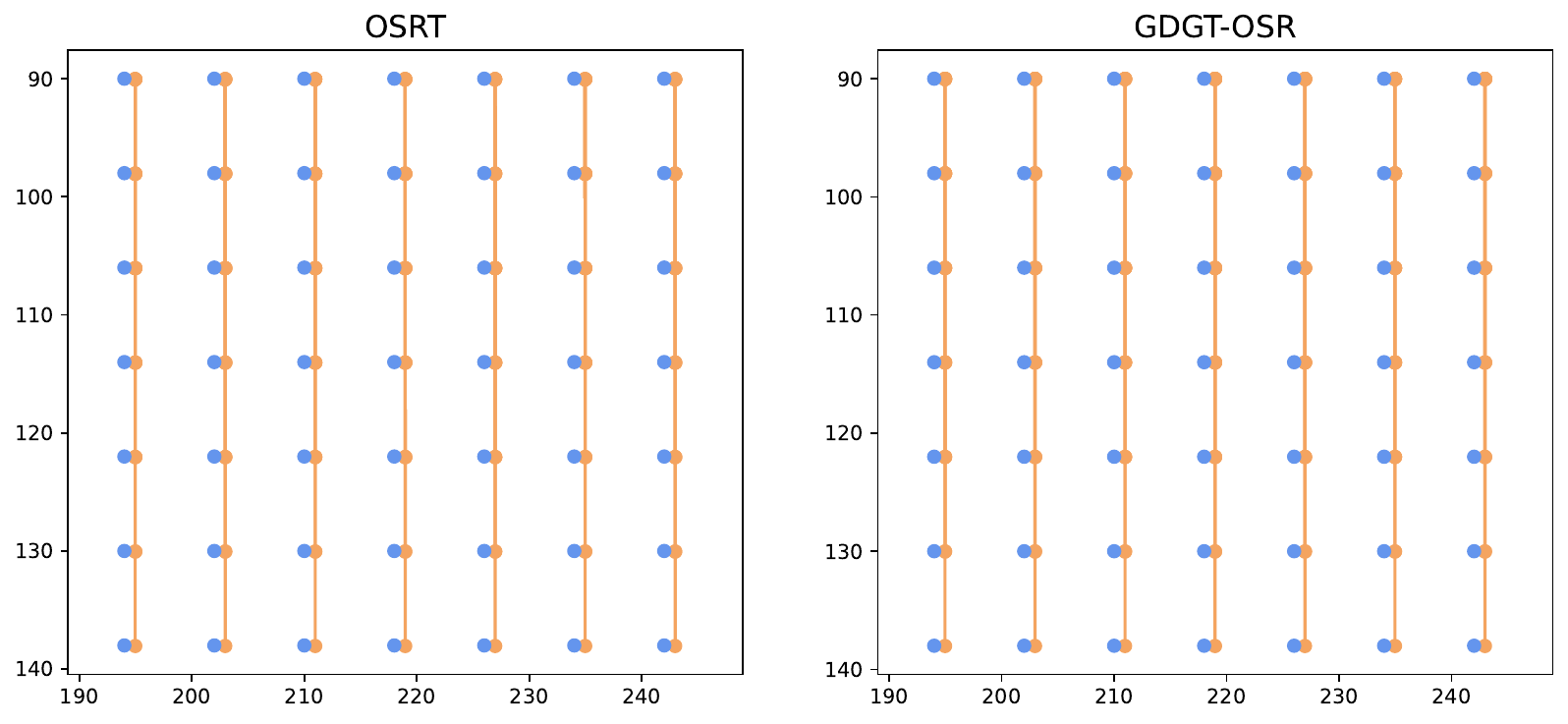}}
\caption{Visualization of offset maps in OSRT \cite{yu2023osrt} and GDGT-OSR. Reference and deformed points are colored in blue and orange. The horizontal and vertical axes denote the x and y coordinates in the image, respectively. (a) Visual comparison of offset maps around the polar area. (b) Visual comparison of offset maps around the equator area.}
\label{SAs}
\end{figure}

\textbf{Impacts of Self-Attentions.} We investigate the influence of DMRSA and DDSA. 
As shown in Table~\ref{MRSA_DDSA}, compared to the variant without both DMRSA and DDSA, the performance of the variant with DMRSA improves significantly. The variant with only DMRSA performs better than that with only DDSA. This illustrates that the proposed DMRSA mechanism dominantly contributes to the ODISR. 
When both DMRSA and DDSA are incorporated into the network architecture, the performance is further improved, which demonstrates the effectiveness of the proposed GDGT-OSR framework. 
Moreover, we investigate the impacts of the offsets in DDSA. Fig.~\ref{SAs} shows the visual comparison of offset maps in DDSAs of both OSRT \cite{yu2023osrt} and GDGT-OSR. We can see that the offsets from our GDGT-OSR are larger than those from OSRT around the polar area where the distortion is the most severe, while the offsets from both OSRT and GDGT-OSR around the equator area are small due to negligible distortion. This observation demonstrates that, when combined with DMRSA, DDSA can adapt better to the distortion and further broaden its attention areas to involve a larger range of pixels. 
The above results and analysis show the effectiveness of collaboration between DMRSA and DDSA.

\begin{figure}[!t]
\centering
\includegraphics[width=0.48\textwidth]{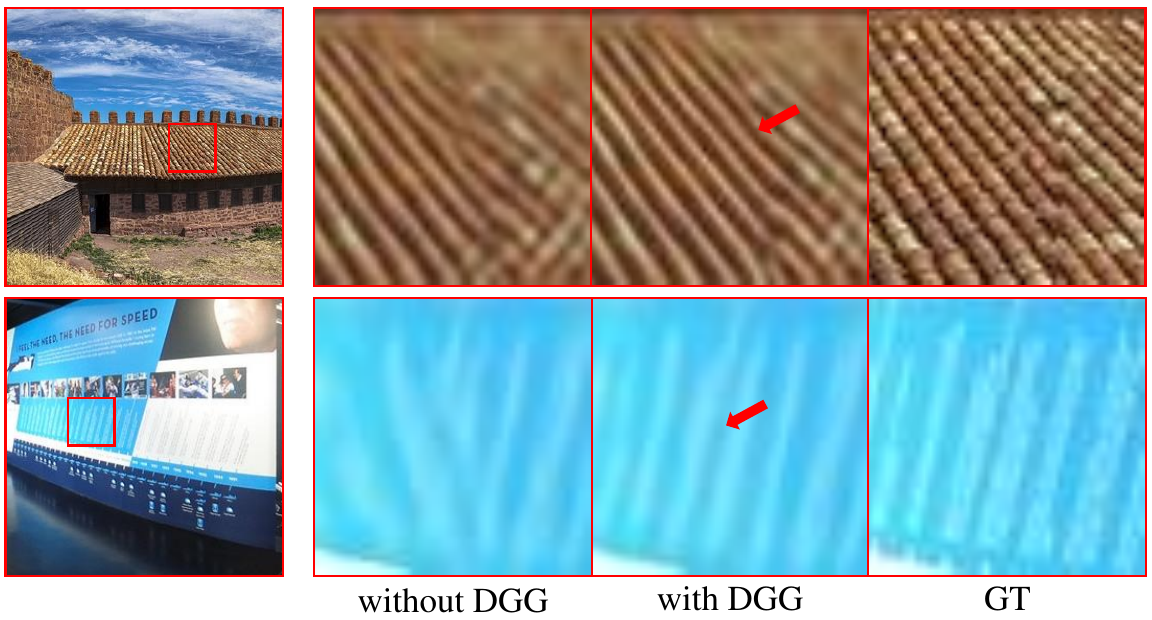}
\caption{Visual comparison of our model without and with DGG. The scaling factor is 4.}
\label{DoM_visual}
\end{figure}

\textbf{Impacts of Distortion Guidance Generator.}
DGG is designed to thoroughly exploit the modulation information of the distortion map for Rwin-SA. As shown in Table~\ref{ablation study}, we present the results of our model without and with DGG in the second and fourth rows under both $\times2$ and $\times4$ scaling factors. With DGG, the model performance surpasses its counterpart (without DGG) across all PSNR-related and SSIM-related metrics, demonstrating the effectiveness of DGG. Fig.~\ref{DoM_visual} shows the qualitative comparison of our model without and with DGG. As shown in the red cropped patches, while the visual results of the model without DGG differ from the ground truth and appear distorted, the model with DGG can restore more visually pleasing results. 
It demonstrates that our model benefits from DGG in enhancing SR performances. 

To further investigate the effects of DGG, we conducted some ablation experiments on the architecture of DGG, as shown in Table~\ref{varients of DoM}, i.e., different ways of exploiting the distortion map. `Direct' represents that the distortion map is replicated along the channel dimension, and then directly utilized to modulate the feature maps. `One conv' means that we use one convolutional layer to replace the proposed DGG. `w/o atten' denotes the DGG without the attention branch. It can be seen that the performance deteriorates most in the `direct' variant. Although we use one convolutional layer to encode the distortion map, the performance is almost the same as that of the `direct' variant. Thus, we can conclude that exploiting the distortion map in simple ways brings little improvement. Moreover, the performance of the `w/o atten' variant is inferior to that of GDGT-OSR, which involves a complete DGG. DGG not only encodes the distortion map into feature maps that match the dimension of the key/value features in Rwin-SA, but also effectively exploits the information of the distortion map.

\begin{figure}[!t]
\centering
\includegraphics[width=0.48\textwidth]{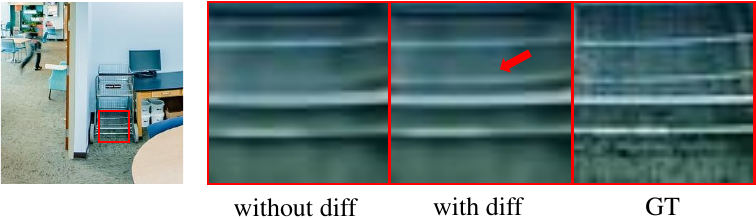}
\caption{Visual comparison of our model without and with the difference calculation in DFA. The scaling factor is 4.}
\label{diff_visual}
\end{figure}

\textbf{Impacts of Diff.}
In DFA, we calculate the difference between $F^{DMRSA}$ and $F^{DDSA}$, which is used to adaptively weigh the combined features. As shown in Table~\ref{ablation study}, the performance slightly degrades without Diff. Fig.~\ref{diff_visual} shows the visual comparison between our model without and with the Diff design in DFA. We can see that some details are missing without Diff, while more details are restored with Diff. It is worth noting that computing the Diff does not require additional parameters, but it can bring improvements in restoring details.
In Table~\ref{ablation study}, we can observe that the SR performance is moderately impacted when both DGG and Diff are absent simultaneously, especially under the $\times2$ scaling factor. This demonstrates the necessity and importance of both utilizing the distortion map and leveraging the distinction between features from DMRSA and DDSA.

\textbf{Impacts of Dynamic Feature Aggregation.}
We propose to aggregate features from DMRSA and DDSA in a dynamic way, based on the learned weights. To verify the efficacy of DFA, we compare the SR performance of different aggregation methods on the SUN 360 panorama dataset, as illustrated in Table~\ref{aggregation}. `Addition' denotes that the two features are summed directly to obtain the final feature. `Concatenation' denotes that the two features are concatenated along the channel dimension, and then convolved by a convolutional layer to reduce the number of channels to match that of the input features. From Table~\ref{aggregation}, we can see that DFA achieves the best or comparative performance across all evaluation metrics compared with other aggregation methods.

\begin{table}[t]
\centering
\caption{Result comparison among different ways of aggregation on SUN 360 PANORAMA \cite{xiao2012recognizing}. The scaling factor is 4.}
\begin{tabular}{c|cccc}
\hline
Ways of Aggregation & PSNR  & SSIM   & WS-PSNR & WS-SSIM \\ \hline\hline
Addition            & 27.88 & \textbf{0.7850} & 27.79   & \textbf{0.8037}  \\ \hline
Concatenation       & 27.89 & 0.7849 & 27.80   & \textbf{0.8037}  \\ \hline
DFA                 & \textbf{27.92} & \textbf{0.7850} & \textbf{27.81}   & 0.8036  \\ \hline
\end{tabular}
\label{aggregation}
\end{table}

\textbf{Impacts of Window Size.}
We investigate the effects of different sizes of rectangle windows, as shown in Table~\ref{window_size}. A window size of (32/8) means that the height and width are 32 and 8 for vertical windows, respectively, and 8 and 32 for horizontal windows. As demonstrated in Table~\ref{window_size}, the network with a window size of (64/8) outperforms those with window sizes of (32/8) and (64/4). This improvement may be because the rectangle window with size (64/8) can effectively capture features of distorted content in a more appropriate range of area.

\textbf{Impacts of Window Shape.}
Besides the window size, the effects of the window shapes of the rectangular windows in DMRSA are also worth exploring. To this end, we conducted experiments using various configurations of rectangular window orientations, i.e., vertical only, horizontal only, and a combination of both vertical and horizontal windows with 50\% each. The results, as presented in Table~\ref{window_shape}, indicate that the configuration with only horizontal rectangular windows outperforms the one with only vertical rectangular windows across both testing datasets. This advantage is likely attributable to the substantial horizontal stretch of projected ERP images, enabling horizontal rectangular windows to capture more self-similar features for enhanced reconstruction. Nonetheless, as demonstrated in Table~\ref{window_shape}, the variant incorporating both vertical and horizontal rectangular windows achieves superior performance. Despite the primary horizontal distortion of ERP images, certain self-similar textures and patterns may still exist vertically, which can be effectively captured using vertical windows. Consequently, our DMRSA mechanism employs a combination of both vertical and horizontal rectangular windows.

\begin{table*}[!t]
\centering
\caption{Ablation study of the window size of DMRSA. The scaling factor is 4.}
\begin{tabular}{c|cccc|cccc}
\hline
\multirow{2}{*}{Window size} & \multicolumn{4}{c|}{ODI-SR\cite{deng2021lau}}        & \multicolumn{4}{c}{SUN 360 Panorama\cite{xiao2012recognizing}}    \\
                             & \multicolumn{1}{c}{PSNR} & SSIM            & \multicolumn{1}{c}{WS-PSNR} & \multicolumn{1}{c|}{WS-SSIM} & \multicolumn{1}{c}{PSNR} & \multicolumn{1}{c}{SSIM} & \multicolumn{1}{c}{WS-PSNR} & \multicolumn{1}{c}{WS-SSIM} \\ \hline\hline
32/8                         & 27.39                    & 0.7768          & 26.67                       & 0.7616                       & 27.82                    & 0.7837                   & 27.73                       & 0.8024                      \\ \hline
64/8                         & \textbf{27.44}           & \textbf{0.7776} & \textbf{26.72}              & \textbf{0.7623}              & \textbf{27.92}           & \textbf{0.7850}          & \textbf{27.81}              & \textbf{0.8036}             \\ \hline
64/4                         & 27.39                    & 0.7746          & 26.67                       & 0.7594                       & 27.81                    & 0.7817                   & 27.71                       & 0.8001                      \\ \hline
\end{tabular}
\label{window_size}
\end{table*}

\begin{table*}[!t]
\centering
\caption{Ablation study of the window shape of DMRSA. The scaling factor is 4.}
\begin{tabular}{c|cccc|cccc}
\hline
\multirow{2}{*}{Window shape} & \multicolumn{4}{c|}{ODI-SR\cite{deng2021lau}}                                                                             & \multicolumn{4}{c}{SUN 360 Panorama\cite{xiao2012recognizing}}                                                                            \\
                             & \multicolumn{1}{c}{PSNR} & SSIM            & \multicolumn{1}{c}{WS-PSNR} & \multicolumn{1}{c|}{WS-SSIM} & \multicolumn{1}{c}{PSNR} & \multicolumn{1}{c}{SSIM} & \multicolumn{1}{c}{WS-PSNR} & \multicolumn{1}{c}{WS-SSIM} \\ \hline\hline
Vertical                         & 27.40                    & 0.7773          & 26.68                       & 0.7619                       & 27.83                    & 0.7842                   & 27.74                       & 0.8027                      \\ \hline
Horizontal                       & 27.42           & 0.7771   & 26.70              & 0.7616              & 27.86           & 0.7842          & 27.76              & 0.8026             \\ \hline
Vertical \& Horizontal           & \textbf{27.44}   & \textbf{0.7776}   & \textbf{26.72}    & \textbf{0.7623}   & \textbf{27.92}    & \textbf{0.7850}   & \textbf{27.81}   & \textbf{0.8036}      \\ \hline
\end{tabular}
\label{window_shape}
\end{table*}

\begin{table*}[!t]
\centering
\caption{Result comparison among different training losses. The scaling factor is 4.}
\begin{tabular}{c|cccc|cccc}
\hline
\multirow{2}{*}{loss} & \multicolumn{4}{c|}{ODI-SR\cite{deng2021lau}}                                          & \multicolumn{4}{c}{SUN 360 Panorama\cite{xiao2012recognizing}}                                \\
                      & PSNR           & SSIM            & WS-PSNR        & WS-SSIM         & PSNR           & SSIM            & WS-PSNR        & WS-SSIM         \\ \hline\hline
$l1$               & 27.43          & \textbf{0.7779} & 26.70          & \textbf{0.7625} & 27.89          & \textbf{0.7854} & 27.78          & \textbf{0.8039} \\ \hline
$l1$-(WS-$l1$)     & 27.41          & 0.7763          & 26.70          & 0.7611          & 27.82          & 0.7834          & 27.75          & 0.8019 \\ \hline
WS-$l1$            & \textbf{27.44} & 0.7776          & \textbf{26.72} & 0.7623          & \textbf{27.92} & 0.7850          & \textbf{27.81} & 0.8036          \\ \hline
\end{tabular}
\label{loss}
\end{table*}

\begin{figure*}[!t]
\centering
\includegraphics[width=\textwidth]{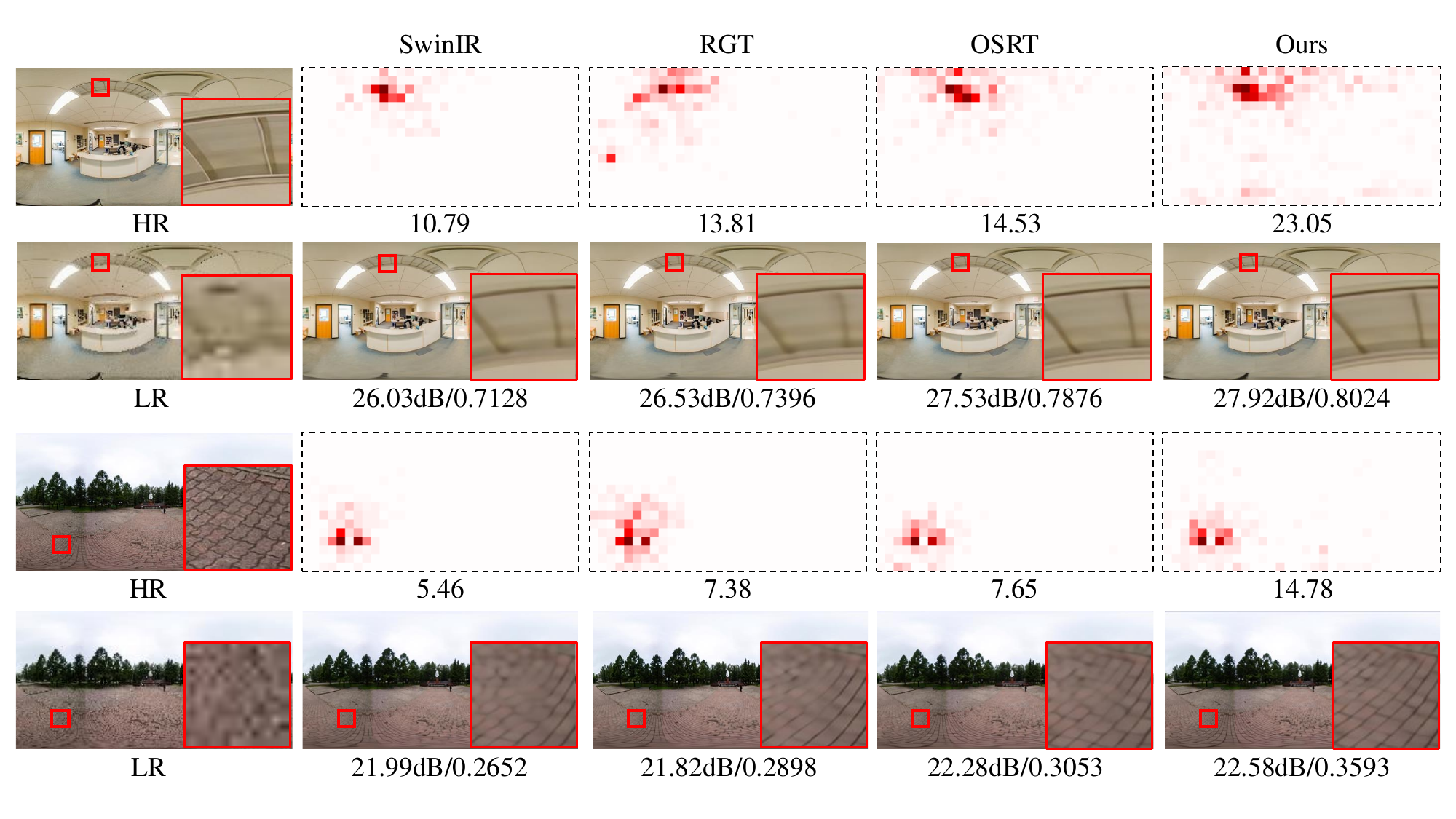}
\caption{Visual comparison of local attribution maps and SR results among different methods. The Diffusion Index (DI) is given below the local attribution maps. A higher DI indicates a wider range of the involved pixels, and vice versa.}
\label{LAM2}
\end{figure*}

\begin{table}[!t]
\centering
\caption{Model complexity comparison. The scaling factor is 8.}
\begin{tabular}{c|c|c|c|c}
\hline
Model & \#Params(M) & \#Multi-Adds(G) & PSNR & WS-PSNR    \\ \hline\hline
SwinIR        & 12.05  & 60.59    & 25.09          & 24.38  \\ \hline
HAT           & 20.92  & 96.05    & 25.15          & 24.42  \\ \hline
RGT           & 13.51  & 62.17    & 24.86          & 24.08  \\ \hline
OSRT          & 12.08  & 54.50    & 25.28          & 24.53  \\ \hline
GDGT-OSR      & 15.99  & 69.74    & 25.32          & 24.60  \\ \hline
\end{tabular}
\label{model_complexity}
\end{table}

\begin{table}[!t]
\centering
\caption{Comparison of average DI values at four different pixels in the SUN 360 Panorama dataset\cite{xiao2012recognizing}.}
\begin{tabular}{ccccc}
\hline
       & (50, 90) & (100, 125) & (150, 200) & (200, 110) \\ \hline\hline
EDSR\cite{lim2017enhanced}   & 1.22   & 1.23     & 1.16     & 1.03     \\ \hline
RCAN\cite{zhang2018image} & 12.29  & 8.77     & 8.09     & 8.25     \\ \hline
SwinIR\cite{liang2021swinir} & 5.63   & 5.09     & 4.88     & 3.97     \\ \hline
OSRT\cite{yu2023osrt}   & 8.30   & 6.88     & 6.84     & 8.61     \\ \hline
GDGT-OSR   & \textbf{16.18}  & \textbf{11.15}    & \textbf{10.37}    & \textbf{19.10}    \\ \hline
\end{tabular}
\label{DI}
\end{table}

\textbf{Impacts of Training Loss.}
Given the non-uniform pixel distribution, we adopt the WS-$l1$ loss as our training loss function to calculate pixel-wise errors in ERP images. From Table~\ref{loss}, we can see that WS-$l1$ outperforms $l1$ in the evaluation metrics of PSNR and WS-PSNR, while $l1$ performs better than WS-$l1$ in the evaluation metrics of SSIM and WS-SSIM. Referred to Eq.~(\ref{ws-l1}), it is probable that in the WS-$l1$ loss, luminance and contrast are affected by the multiplication of the distortion map in the training process, which leads to the degradation of SSIM and WS-SSIM. However, the performance of PSNR and WS-PSNR demonstrates that WS-$l1$ loss can improve and relieve pixel-wise errors.

Eq.~(\ref{ws-l1}) indicates that the WS-$l1$ loss is obtained by weighing the $l1$ loss with the distortion map $D$. To further study the effectiveness and rationality of WS-$l1$ loss as the training loss, we train the model with a loss that weighs the $l1$ loss with (1-$D$). This loss assigns larger weights to the high-latitude areas and smaller weights to the low-latitude areas, which is opposite to the WS-$l1$ loss. 
It can easily be written as $l1-(\text{WS-}l1)$. As shown in Table~\ref{loss}, the model trained with $l1-(\text{WS-}l1)$ yields inferior results compared to those trained with $l1$ and WS-$l1$. These findings suggest that prioritizing lower-latitude regions, where pixels are more densely distributed, is crucial during the training process of ODISR.

\subsection{Model Complexity}
We compare the model complexity of our methods with other SOTA methods, as summarized in Table~\ref{model_complexity}. We evaluate the model performance on the ODI-SR dataset under the $\times8$ scaling factor. The number of Multi-Adds is calculated for an input size of $64\times64$. As shown in Table~\ref{model_complexity}, the number of parameters and Multi-Adds of HAT\cite{chen2023hat} are much larger than those of other methods. However, despite its substantial computational complexity, the performance of HAT\cite{chen2023hat} remains inferior to both OSRT\cite{yu2023osrt} and our GDGT-OSR. On the contrary, compared to other methods, GDGT-OSR achieves significant performance improvements with a comparable or only marginally increased number of parameters. This highlights the superiority and effectiveness of our GDGT-OSR.

\subsection{Exploration on Involved Area.} To investigate the effects of the range of involved pixels, we randomly selected four different pixels from each image in the SUN 360 Panorama testing dataset and calculated their average DI values, as shown in Table~\ref{DI}. Higher DI values indicate a wider range of pixels involved. Note that the coordinates of the upper left corner of an image are (0,0). As shown in Table~\ref{DI}, the proposed method achieves the highest DI values at the four selected pixels from different latitudes compared to other methods, which demonstrates that our method has the widest range of involved area. Fig.~\ref{sample} visually shows that our method's area of contribution for reconstructing the patch in the red box is the widest with more self-similar textures, resulting in superior SR performance. 
More visual results of local attribution maps are illustrated in Fig.~\ref{LAM2}, demonstrating that the proposed GDGT-OSR method effectively leverages a broader range of pixels. This capability enables it to produce more visually pleasing and higher-quality SR results.
The above results indicate that the proposed mechanisms can enlarge the attention area and improve SR performance. 

\section{Summary}

\subsection{Limitations}
\textbf{Inference Speed.}
Typically, ODI images possess high resolutions, such as 2K, 4K, or even 8K. However, the inference speed of the proposed method is limited when processing these high-resolution ODI images due to the computationally intensive self-attention mechanisms employed. Furthermore, during inference, if the model size is excessively large, GPU memory may become insufficient to handle such high-resolution inputs. Although GDGT-OSR can produce high-quality SR results, its practical application is constrained in scenarios requiring real-time super-resolution of high-resolution ODI images with limited computational resources.

\begin{figure}[!t]
\centering
\includegraphics[width=0.49\textwidth]{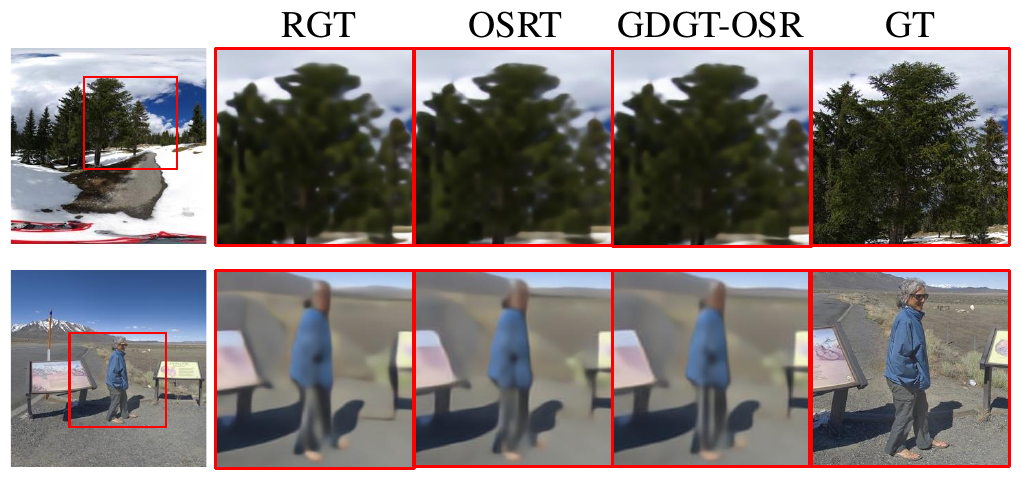}
\caption{Failed cases under the $\times16$ scaling factor.}
\label{failed_case}
\end{figure}
\textbf{Large Scaling Factors.}
When super-resolving objects with numerous high-frequency details, e.g., trees and humans, under the large scaling factor of $\times16$, the qualitative results produced by GDGT-OSR are not visually satisfactory, as illustrated in Fig.~\ref{failed_case}. The primary reason is the significant loss of fine-grained information in the LR images when using a large scaling factor. With limited available information in the LR images, it becomes challenging for GDGT-OSR, as well as other SOTA methods, such as RGT\cite{chen2024recursive} and OSRT\cite{yu2023osrt}, to reconstruct visually pleasing results for these high-frequency components in such demanding conditions.

\subsection{Conclusion}
\label{sec:conclusion}
In this paper, we propose a Geometric Distortion Guided Transformer for Omnidirectional image Super-Resolution, named GDGT-OSR. Considering the geometric distortion in ERP images, we propose a Distortion Modulated Rectangle-window Self-Attention (DMRSA) mechanism, integrated with Distortion-aware Deformable Self-Attention (DDSA), to adapt to the unevenly distorted content. In this way, GDGT-OSR captures features from the attention areas with various shapes, aiming to calibrate the attention regions and facilitate their expansion, capturing more self-similar and related textures. To exploit the distortion map, we propose a Distortion Guidance Generator (DGG) to transform geometric distortion into distortion guidance, which is leveraged to modulate the key and value features in Rwin-SA. Furthermore, we adaptively aggregate two features generated by DMRSA and DDSA through a Dynamic Feature Aggregation (DFA) module. Experiment results demonstrate that GDGT-OSR can restore more details and richer textures over other methods, achieving SOTA performance on omnidirectional image super-resolution.




\bibliographystyle{IEEEtran}
\bibliography{refs}

\end{document}